\documentclass[apj]{emulateapj}

\usepackage{amssymb}
\usepackage{amsmath}
\usepackage{latexsym}
\usepackage{natbib}
\usepackage{wasysym}
\usepackage{mathtools}
\usepackage[hang,flushmargin]{footmisc} 

\DeclareMathAlphabet{\mathpzc}{OT1}{pzc}{m}{it}

\newcommand\be{\begin{equation}}
\newcommand\ee{\end{equation}}

\begin{document}

\pagenumbering{arabic}

\shorttitle{Resonant Chains}
\shortauthors{MacDonald \& Dawson}

\title{Three Pathways for Observed Resonant Chains}
\author{Mariah G. MacDonald\altaffilmark{1}, Rebekah I. Dawson\altaffilmark{1,2}}
\email{mmacdonald@psu.edu}

\altaffiltext{1}{Department of Astronomy \& Astrophysics, The Pennsylvania State University}
\altaffiltext{2}{Center for Exoplanets and Habitable Worlds, The Pennsylvania State University}

\setcounter{footnote}{0}

\begin{abstract} 
A question driving many studies is whether the thousands of exoplanets known today typically formed where we observe them or formed further out in the disk and migrated in. Early discoveries of giant exoplanets orbiting near their host stars and exoplanets in or near mean motion resonances were interpreted as evidence for migration and its crucial role in the beginnings of planetary systems. long-scale migration has been invoked to explain systems of planets in mean motion resonant chains consisting of three or more planets linked by integer period ratios. However, recent studies have reproduced specific resonant chains in systems via short-scale migration, and eccentricity damping has been shown to capture planets into resonant chains. We investigate whether the observed resonant chains in Kepler-80, Kepler-223, Kepler-60, and TRAPPIST-1 can be established through long-scale migration, short-scale migration, and/or only eccentricity damping by running suites of N-body simulations. We find that, for each system, all three mechanisms are able to reproduce the observed resonant chains. long-scale migration is not the only plausible explanation for resonant chains in these systems, and resonant chains are potentially compatible with in situ formation.
\end{abstract}

\keywords{planets and satellites: dynamical evolution and stability; stars: individual (Kepler-80, Kepler-223, Kepler-60, TRAPPIST-1)}



\section{Introduction}
\setcounter{footnote}{0}

An open question is whether the thousands of exoplanets detected to date formed where we observe them today or whether migration has altered their orbital properties. Early discoveries of giant exoplanets orbiting near their host stars \citep[e.g., 51 Peg b,][]{Mayor1995} and exoplanets in/near mean motion resonances \citep[e.g., GJ 876 b/c,][]{Marcy2001} were interpreted as evidence for gas disk migration and its crucial role in the beginnings of planetary systems \citep[e.g.,][]{Lin1996,Rasio1996,Lee2002}. Migration is considered a robust physical process as any asymmetry in disk conditions inside vs. outside the planet's orbit will result in a net torque that drives migration.

Given the theoretical robustness of gas disk migration and its invoked effect on a few observed systems, many studies predicting the properties of planets have assumed that migration is key in establishing the orbits, compositions, and occurrence rates of planets within a few AU of their star \citep[e.g.,][]{Ida2008,Mordasini2009,Rein2012b,Cossou2014}. However, even though gas disk migration is able to explain the origin of individual systems and has been assumed to be important in establishing the properties of the observed planetary population, recent observational and theoretical studies have called into question the prevalence of planet gas disk migration. New studies have found that close-in planets might have formed in situ \citep{Chiang2013,Lee2014,Batygin2016,Lee2017}, and others have shown that properties of the $Kepler$ population (e.g., spacings, mass, compositions) can be reproduced without invoking migration \citep[e.g.,][MacDonald \& Dawson, in prep.]{Hansen2012,Hansen2013,Chiang2013,Lee2014,Dawson2015,Dawson2016,Moriarty2016}. In addition, migration is expected to drive planets into orbital resonance (e.g., \citep{Malhotra1993,Melita1996,Lee2002,Terquem2007,Mustill2011,Wang2014}), but most systems discovered by $Kepler$ are not in or near orbital resonances \citep{Lissauer2011,Veras2012,Fabrycky2014}.

Migration has been invoked to explain systems of bodies in mean motion resonant (MMR) chains, most recently for exoplanet systems for which the observations constrain resonant angles \citep[e.g.,][]{Mills2016,MacDonald2016}. Several of these systems feature resonant chains, which consist of three or more planets in a system linked by integer period ratios. For example, Jupiter's large moons Io, Ganymede, and Europa have orbital period ratios of 1:2:4, a configuration thought to have originated during the satellites’ circumplanetary disk-driven \citep[][]{Peale2002} or tidal \citep[][]{Yoder1979} migration. Some resonant chains consist of a series of two-body mean motion resonances, with each resonant angle involving orbital angles of two planets. Two-body resonant angles take the form
\begin{equation}
    \Theta_{b-c} = j_1\lambda_b + j_2\lambda_c + j_3\omega_b + j_4\omega_c + j_5\Omega_b+j_6\Omega_c,
\end{equation}
\noindent where $b$ and $c$ refer to two planets, $\lambda$ is the mean longitude, $\omega$ is the longitude of periapse, $\Omega$ is the longitude of ascending node, and the $j$ coefficients must sum to zero.  Some resonant chains instead or additionally feature three-body MMR, with resonant angles involving orbital angles of three planets.  Three-body resonant angles take the form
\begin{equation}
\label{eqn:three}
   \phi = p \lambda_1 - (p+q) \lambda_2 + q \lambda_3,
\end{equation}
\noindent where $\lambda_i$ is the mean longitude of the $i$th body and $p$ and $q$ are integers. The commensurability in periods creates a repeating geometrical configuration of three-planets. Eqn. \ref{eqn:three} is for zeroth-order three-body resonances, which are by far the strongest in the case of small eccentricities \citep{Gallardo2016}. When dynamical interactions cause the resonant angle $\phi$ to librate, we consider this commensurability to be a three-body resonance. Certain combinations of librating two-body angles guarantee libration of a three-body angle. A system might instead have libration of multiple two-body resonant angles without libration of three-body resonant angles. We investigate both types of resonant chain configurations as well as three-body resonances without two-body resonances. 

Resonant chains are often considered hallmarks of planetary migration. For example, \citet{Mills2016} argued that the architecture of Kepler-223 requires gas disk migration. In their proof of concept migration simulations, the planets start beyond 1 AU and migrate to $\sim$ 0.1 AU star separations. However, other studies have shown that capture into resonant chains can sometimes occur without this long-scale migration. In simulations in which the planets' periods decrease by only a few percent, \citet{MacDonald2016} captured the Kepler-80 planets into their resonant chain. \citet{Dong2016} found that systems of giant planets can be captured into resonant chains without migration in simulations that included only eccentricity damping.

In this work, we assess the  necessity of migration in known resonant chain systems. We investigate whether the known or suspected resonant chains in Kepler-80, Kepler-223, Kepler-60, and TRAPPIST-1 can be established through this long-scale migration, as well as through short-scale migration and eccentricity damping. For each system, a resonant chain has been established in the literature in a simulation through one mechanism as a proof of concept, but it has not yet been studied whether any other mechanism is effective. These systems are all of the super-Earth systems with resonant chains. With these three dynamical histories (long-scale migration, short-scale migration, eccentricity damping only), we aim to assess if long-scale migration is integral to the formation of resonant chains, or if other mechanisms are plausible.

In Section \ref{sec:simulations}, we describe our simulations and how the three dynamical histories are implemented. We analyze our simulations for each system in Sections \ref{sec:K80}--\ref{sec:trap1}. We discuss these results and conclude in Section \ref{sec:conclusion}.


\section{Simulations}\label{sec:simulations}

In an attempt to reproduce the resonant chains known or suspected in Kepler-80, Kepler-223,  Kepler-60, and TRAPPIST-1, we simulate each system under three dynamical histories: long-scale migration, short-scale migration, and eccentricity damping only.  In long-scale migration, the planets start far beyond their present day orbits, with the inner-most planet starting at 1 AU and the other planets starting wide of their observed period ratios. In short-scale migration, the planets start just wide of their observed periods, essentially in situ, and only migrate for a short distance until they reach their observed periods and lock into the appropriate resonances.

For each system and dynamical history, we run a large number of N-body simulations with all planet-pair period ratios initialized wide of their observed commensurabilities.  The masses and sky-plane inclinations were drawn from independent normal distributions based on the values reported by previous studies. The longitudes of ascending node and arguments of periapse were drawn from uniform distributions. Initial eccentricities for all planets were set to $e=0.0$, except for simulations with eccentricity damping only (see Table \ref{tab:simIC}). All integrations were performed using the \verb|WHFAST| integrator \citep{Rein2015} in the open-source \verb|REBOUND| N-body package \citep{Rein2012}. We adopt a timestep of 5\% of the innermost planet's observed orbital period. 

We applied an inward migration force and/or eccentricity damping forces (at constant angular momentum for the latter) on timescales  $\tau_a \sim 10^6-10^{11}$  and $\tau_e \sim 10^{4}-10^7$, respectively, following the prescription in \citet{Papaloizou2000}. These timescales were drawn from independent log-uniform distributions. For migration, whether short-scale or long-scale, these forces were applied only to the outermost planet in the system\footnote{We do not know the  migration rate or direction of migration for each planet since they depend on the conditions of the disk. By simulating only the outer planet migrating, we implicitly assume the case where the migration of the inner planets is on a much longer timescale. Future work could explore how the relative migration rates of the planets affect their final resonant configurations.}. In simulations that included eccentricity damping only, the damping was applied to all planets. The migration and eccentricity damping forces used as described above were applied using the \verb|modify_orbits_forces| routine in the REBOUNDx library.  These forces were turned off after integrating the system for $10^6$ -- $10^8$ days; the system was then integrated forward for an additional $10^7$ -- $10^8$ days to assess stability. We deemed a system unstable and stopped the simulation if any of the planets reached $e < 0$, $e \geq 1.0$, or  $P< 0,$ where $e$ and $P$ are the planet's eccentricity and period, respectively. The parameters used for each suite of simulations, including $\tau_a$, $\tau_e$, the length of time migration forces were applied, and the total length of the simulation, are listed in Tables \ref{tab:simIC} and \ref{tab:simIC2}.

For each simulation, we plot and examine by eye each associated three-body angles and two-body angles to determine whether the angle is librating. We list the number of simulated systems that survived, i.e., did not reach any of the instability criteria listed above, and the number of systems in resonance by the end of the simulation in Table \ref{tab:simnum}. A system was deemed to be in resonance if all of the two-body angles were librating and/or if all of the three-body angles were librating.

\begin{deluxetable*}{rcccccc}
\tablecaption{Simulation Initial Conditions\label{tab:simIC}}
\tablecolumns{7}
\tablewidth{0pt}
\tabletypesize{\scriptsize}
\tablehead{\colhead{}&\colhead{}&\colhead{}&\colhead{}&\colhead{}&\colhead{}&\colhead{}}   

\startdata

Kepler-80 & d & e & b & c  &  & \\
\hline\\
P (d) & 3.072 & 4.645 & 7.052 & 9.524  & &  \\
m ($M_{\oplus}$) & N(6.48,0.46) & N(4.92,0.49) & N(5.99,0.57) & N(5.03,0.42)  & &  \\
$i$ (deg) & N(88.35,1.51) & N(88.79,1.07) & N(89.34,0.62) & N(89.33,0.57)  &  & \\ 
$e$ & 0.0 & 0.0 & 0.0 & 0.0   &  \\

P$_{sm}^i$(d) & \multicolumn{4}{c}{[N(0.09,0.03)+1]$\times$ P}  &  &   \\
P$_{lm}^i$(d) & 427.5 & 648.3 & 989.7 & 1395.0 & \\
P$_{ecc}^i$(d) & \multicolumn{4}{c}{[N(0.09,0.03)+1]$\times$ P}  &  &  \\
$e_{ecc}^i$ & 0.04 & 0.04 & 0.04 & 0.04  &  &  \\ 
 
\cline{2-7} \\
mech   & No. run   & $\tau_a$(d)          & $\tau_e$(d)       & $T_1$ (d)  &  $T_2$ (d)  &  \\
sm & 300   & $10^6-10^8$ & $10^3-10^6$& $5\times10^6$  & $9\times10^7$ &  \\
lm & 300& $10^8-10^{11}$ &$10^4-10^6$& $\tau_a*7.5$ & $T_1*3$ & \\
ecc& 300  & \nodata        & $10^3-10^6$   & N(5,1)$\times10^6$ & $T_1*3$  & \\
\hline \\
Kepler-223 & b & c & d & e & &  \\
\hline\\
P (d) & 7.384 & 9.846 & 14.789 & 19.726   & & \\
m ($M_{\oplus}$) & N(7.4,1.3) & N(5.1,1.7) & N(8.0,1.5) & N(4.8,1.4) &  & \\
inc (deg) & N(90.0,1.8) & N(90.0,1.3) & N(87.94,0.32) & N(88.00,0.27)  &  & \\
ecc & 0.078 & 0.15 & 0.037 & 0.051 &  &  \\

P$_{sm}^i$(d) & \multicolumn{4}{c}{[N(0.09,0.03)+1]$\times$ P} &  & \\
P$_{lm}^i$(d) & 344.4 & 473.2 & 751.2 & 1129.4 &   & \\
P$_{ecc}^i$(d) & \multicolumn{4}{c}{[N(0.09,0.03)+1]$\times$ P} & &  \\
$e_{ecc}^i$ & N(0.078,0.01) & N(0.15,0.005) & N(0.037,0.005) & N(0.051,0.005)&  &  \\ 

 \cline{2-7} \\
 mech   & No. run   & $\tau_a$(d)          & $\tau_e$(d)         & $T_1$ (d)  &  $T_2$ (d)    &   \\
 sm     & 300  & $10^6-10^8$ & $10^3-10^6$   & $5\times10^6$ &  $9\times10^7$ &  \\
 lm     & 300   & $10^8-10^{11}$   & $10^4-10^7$   & $5\times10^10$   & $T_1*3$  & \\
 ecc    & 300       & \nodata  & $10^3-10^6$   & $9\times10^7$ &  $9\times10^7$ &  \\
 \hline \\
Kepler-60 & b & c & d &  &  & \\
\hline\\
P (d) & 7.1334 & 8.9187 & 11.8981 &  & &   \\
m ($M_{\oplus}$) & N(4.42,0.97) & N(4.09,1.59) & N(4.42,0.81) &   &  & \\
inc (deg) & 90.0 & 90.0 & 90.0 &   & &  \\
ecc & 0.0 & 0.0 & 0.0 &  & & \\

P$_{sm}^i$(d) & \multicolumn{3}{c}{[N(0.09,0.03)+1]$\times$ P} & & &  \\
P$_{lm}^i$(d) & 550.6 & 730.7 & 1000.0 & & \\
P$_{ecc}^i$(d) & \multicolumn{3}{c}{[N(0.09,0.03)+1]$\times$ P} & & &  \\
$e_{ecc}^i$ & 0.05 & 0.05 & 0.05 &   & &  \\ 

 \cline{2-7} \\
mech    & No. run   & $\tau_a$(d)       & $\tau_e$(d)  & $T_1$ (d)  &  $T_2$ (d)  &  \\
sm      & 300       & $10^6-10^8$       & $10^3-10^6$  & $5\times10^6$ & $9\times10^7$     &   \\
lm      & 300       & $10^8-10^{10}$    & $10^4-10^6$   & $4\times \tau_a$ & $T_1*3$ &   \\
ecc     & 300       & \nodata           & $10^3-10^6$   & $9\times10^7$  & $9\times10^7$  &  \\

\enddata
\tablecomments{Summary of the initial conditions used for the simulations, including the initial masses (m), sky-plane inclinations ($i$), periods (P$^i$), and eccentricities ($e^i$) for each planet, as well as the semi-major axis damping timescale ($\tau_a$), eccentricity damping timescale ($\tau_e$),  the length of the damping before it was turned off ($T_1$), and the total length of the simulation ($T_2$). The two evolution timescales were drawn from log uniform distributions spanning the range given. The table also includes the present day periods of each of the planets, since many of the starting periods depend on them. Here, `No. runs' stands for the number of simulations run for each dynamical history. `sm', `lm' and `ecc' stand for short-scale migration, long-scale migration, and eccentricity damping, respectively. Ref -- Kepler-80: \citet{MacDonald2016}, Kepler-223: \citet{Mills2016}, Kepler-60: \citet{Jontof-Hutter2016}. }
\end{deluxetable*}

\begin{deluxetable*}{rcccccc}
\tablecaption{Simulation Initial Conditions\label{tab:simIC2}}
\tablecolumns{7}
\tablewidth{0pt}
\tabletypesize{\footnotesize}
\tablehead{\colhead{}&\colhead{}&\colhead{}&\colhead{}&\colhead{}&\colhead{}&\colhead{}}   

\startdata
TRAPPIST-1 & b & c & d & e  & f & g \\
\hline\\
P (d) & 1.511 & 2.422 & 4.050 & 6.100  & 9.207 & 12.353 \\
m ($M_{\oplus}$) & N(0.85,0.72) & N(1.38,0.61) & N(0.41,0.27) & N(0.62,0.58)  & N(0.68,0.18) & N(1.34,0.88) \\
$i$ (deg) & N(89.65,0.27) & N(89.67,0.17) & N(89.75,0.16) & N(89.86,0.12)  & N(89.68,0.034) & N(89.71,0.023) \\ 
$e$ & 0.0 & 0.0 & 0.0 & 0.0  &  0.0 &  0.0\\

P$_{sm}^i$(d) & \multicolumn{6}{c}{[N(0.09,0.03)+1]$\times$ P}  \\
P$_{lm}^i$(d) & 1291 & 2324 & 3952 & 6322 & 10116 & 14162 \\
P$_{ecc}^i$(d) & \multicolumn{6}{c}{[N(0.05,0.03)+1]$\times$ P} \\
$e_{ecc}^i$ & 0.04 & 0.04 & 0.04 & 0.04  &0.04  & 0.04 \\ 
 
\cline{2-7} \\
mech   & No. run   & $\tau_a$(d)          & $\tau_e$(d)       & $T_1$ (d)  &  $T_2$ (d) & \\
sm & 500    & $10^6-10^8$ & $10^3-10^6$& $5\times10^6$  & $9\times10^7$  & \\
lm & 600    & $10^7-10^{9}$ &$10^5-10^7$& $\tau_a*7$ & $T_1*3$ & \\
ecc& 600    & \nodata        & $10^3-10^6$   & N(5,1)$\times10^6$ & $T_1*3$ & 
\enddata
\tablecomments{Summary of the initial conditions used for the simulations. `sm', `lm' and `ecc' stand for short-scale migration, long-scale migration, and eccentricity damping, respectively. For definition of other parameters, see Table \ref{tab:simIC}. Ref --  \citet{Tamayo2017}. }
\end{deluxetable*}

\begin{deluxetable}{rccc}
\tablecaption{Simulation Outcomes\label{tab:simnum}}
\tablecolumns{4}
\tablewidth{0pt}
\tabletypesize{\footnotesize}
\tablehead{\colhead{mech}&\colhead{No. run}&\colhead{No. sur}&\colhead{No. res}}   

\startdata

Kepler-80 &  &  &  \\
lm &    300  &  82  (0.27)  & 80  (0.98)\\
sm &    300  &  262 (0.87)  & 257 (0.98) \\
ecc&    500  &  442 (0.88)  & 135 (0.31)\\
 \\

Kepler-223 &  &  &  \\
 lm     & 300   &  51   (0.17)  &  28  (0.55) \\
 sm     & 300  &  240   (0.8)  &  205  (0.85) \\
 ecc    & 300   & 137   (0.46)  &  19 (0.14) \\ 
 \\
 
Kepler-60 & & &  \\
lm      & 300    & 209 (0.70)   & 72  (0.34) \\ 
sm      & 300    & 239 (0.80)   & 85  (0.36) \\
ecc     & 300    & 262 (0.87)   & 146 (0.56) \\
\\ 

TRAPPIST-1 & & &  \\
lm      & 600    & 5  (0.008)   & 4  (0.80) \\ 
sm      & 300    & 13 (0.043)   & 11 (0.85) \\
ecc     & 300    & 34 (0.11)   & 5 (0.15)
\enddata
\tablecomments{Summary of the results of the simulations. `No. runs' stands for the number of simulations run for each dynamical history, `No. sur' indicates the number of systems that did not go unstable (as described by the instability criteria in Section \ref{sec:simulations}), and `No. res' indicates the number of stable systems that were in resonance by the end of the simulation. `lm', `sm' and `ecc' stand for long-scale migration, short-scale migration, and eccentricity damping, respectively. The parenthetical number is the fraction of systems matching those criteria.}
\end{deluxetable}



\section{Kepler-80}\label{sec:K80}

Kepler-80 is a K dwarf hosting a system of six planets, with orbital periods ranging from 0.99 to 14.6 days and planetary masses ranging from ~1--7 $M_{\oplus}$. The planet ordering in orbital period is: f, d, e, b, c, g. The planets b and c were confirmed in 2013 by anti-correlated TTVs \citep[][]{Xie2013}, planets d and e were validated by \citet{Lissauer2014} and \citet{Rowe2014}, the innermost prilanet f was statistically validated by \citet{Morton2016}, and the outermost planet g was most recently statistically validated via neural network \citep{Shallue2018}. Masses and orbits for the middle four planets (d, e, b, c) were measured via TTV fitting \citep[]{MacDonald2016}. These four planets are dynamically intertwined, with four three-body angles librating. \citet{MacDonald2016} showed that the planets could lock into resonance if they migrated from only a few percent beyond their observed periods.

\citet{MacDonald2016} measured the three-body angles for this system. They found the three body resonance angle among the outer three planets (e, b, and c)  $\phi_1=2\lambda_e-3\lambda_b+1\lambda_c$ to librate around $-162^{\circ}$ with an amplitude of $\sim1.0^{\circ}$. The angle $\phi_2=3\lambda_d-5\lambda_e+2\lambda_b$ among the inner three planets (d, e, and b) also librates with a center of $-72.5^{\circ}$ and amplitude of $\sim1.4^{\circ}$. \citet{MacDonald2016}  simulated short-scale migration for the system and reproduced these centers and amplitudes. The other two three-body angles that were found to be librating are a combination of planets that are not all adjacent (i.e. angles involving (d,e,c) and (d,b,c)). In their TTV analysis of the system, \citet{MacDonald2016} limited the eccentricities to $e\cos\omega<0.02$ and $e\sin\omega<0.02$, so their fitted eccentricities are constrained by this prior.  

We aim to reproduce the observed resonant chain by simulating the planets through long-scale migration, short-scale migration, and eccentricity damping. In these simulations, we additionally aim to reproduce period ratios between adjacent planets that are consistent with the measured values as well as eccentricities that are below the upper limit imposed by \citet{MacDonald2016}. For all three suites of simulations, we assume a stellar mass of $M=0.74M_{\odot}$ \citep{MacDonald2016}. We draw values for planetary masses and inclinations  from normal distributions centered on values from \citet{MacDonald2016}, and set all eccentricities for the short and long-scale migration simulations to zero. For the eccentricity damping simulations, all eccentricities are initialized at 0.04. All other values are set as described in Section \ref{sec:simulations}. \citet{MacDonald2016} found that the innermost planet f did not dynamically interact with the outer four planets locked in the resonance chain, and therefore we do not include it in our simulations. We also do not model planet g\footnote{The validation of outermost planet g occurred following the completion of this work. It is most likely part of this resonant chain, but is not modeled in this study.} as it was not included in previous studies which we are comparing to.

\begin{figure*}[t]
    \centering
    \includegraphics[width=0.33\textwidth]{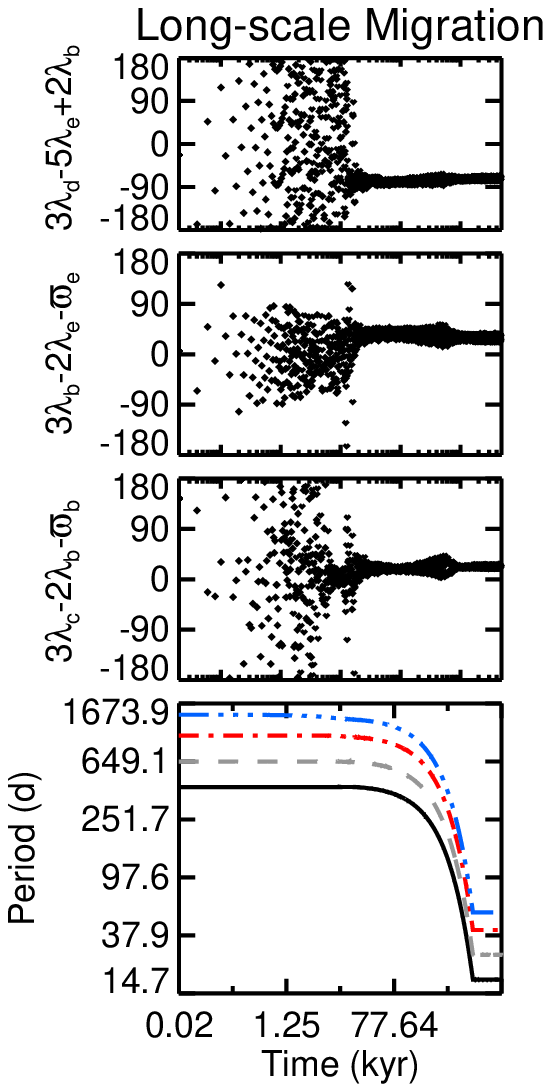}
    \includegraphics[width=0.33\textwidth]{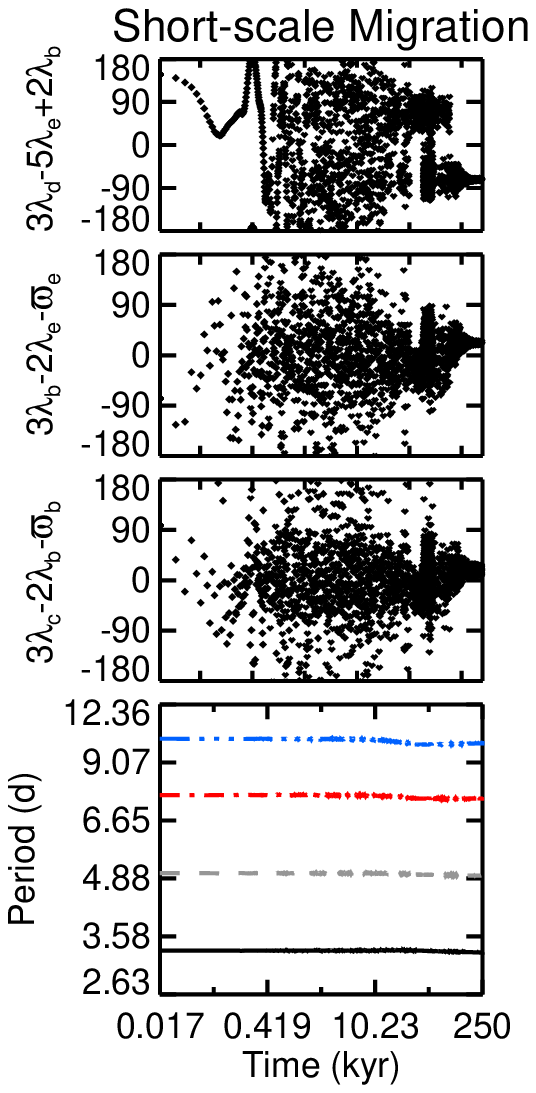}
    \includegraphics[width=0.33\textwidth]{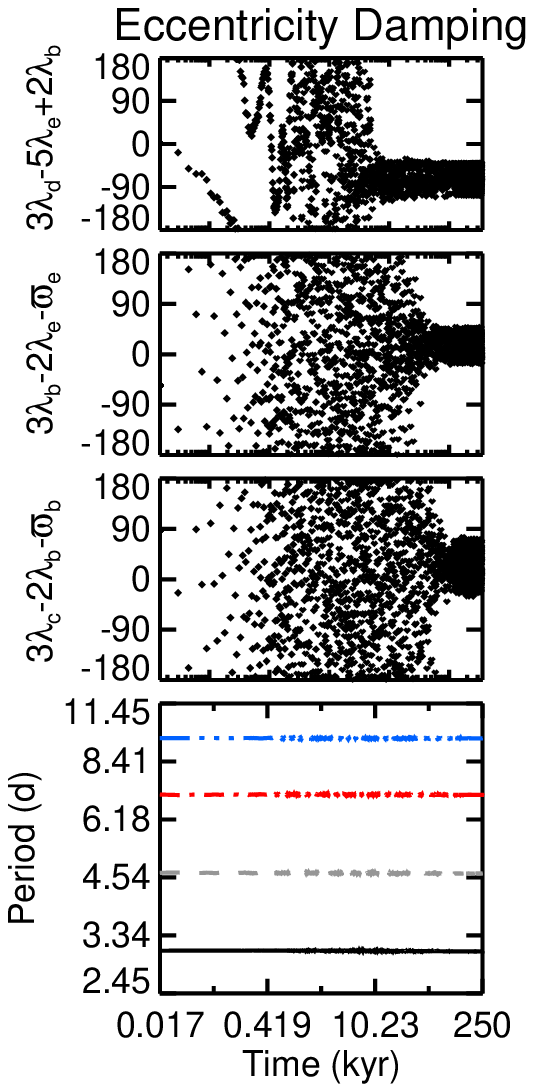}
    \caption{Simulations of each dynamical history can produce the resonant configuration observed in Kepler-80. Panels from left to right: reproduced via long-scale migration, short-scale migration, and eccentricity damping. Each panel shows the three-body angle $\phi_1$, the two-body angles associated with the three-body angle, and the planets' periods. Note that although only one of the three-body angles is shown here, as well as only two of the various two-body angles, they are all  librating. Evolution timescales in days for long-scale migration, short-scale migration, and eccentricity damping: $\tau_a=7.8\times10^7 $, $\tau_e =2.3\times10^5 $, $\tau_a=1.3\times10^8 $, $\tau_e = 1.2\times10^6$, $\tau_e = 4.0\times10^5$.}
    \label{fig:k80}
\end{figure*}

\begin{figure*}[h]
    \centering
    \includegraphics[width=0.3\textwidth]{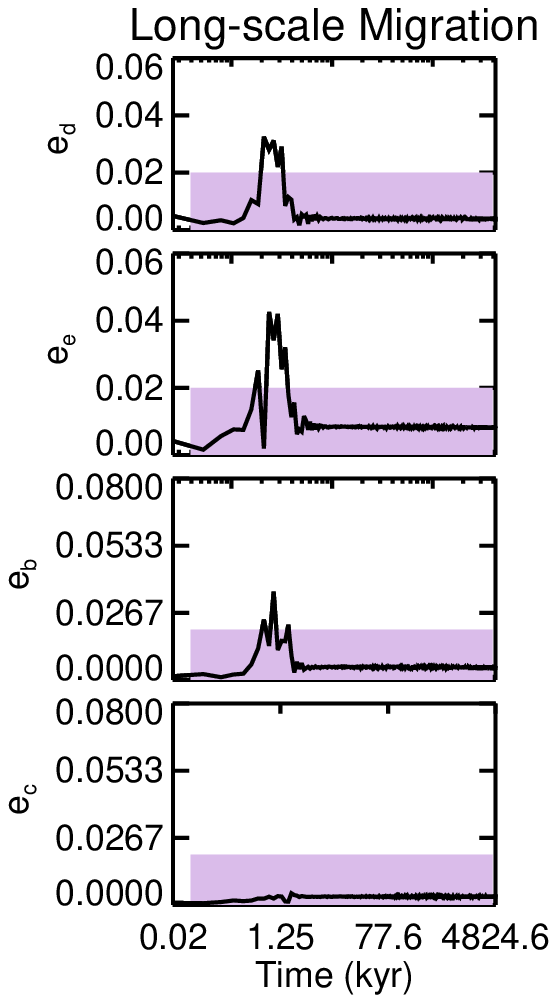}
    \includegraphics[width=0.3\textwidth]{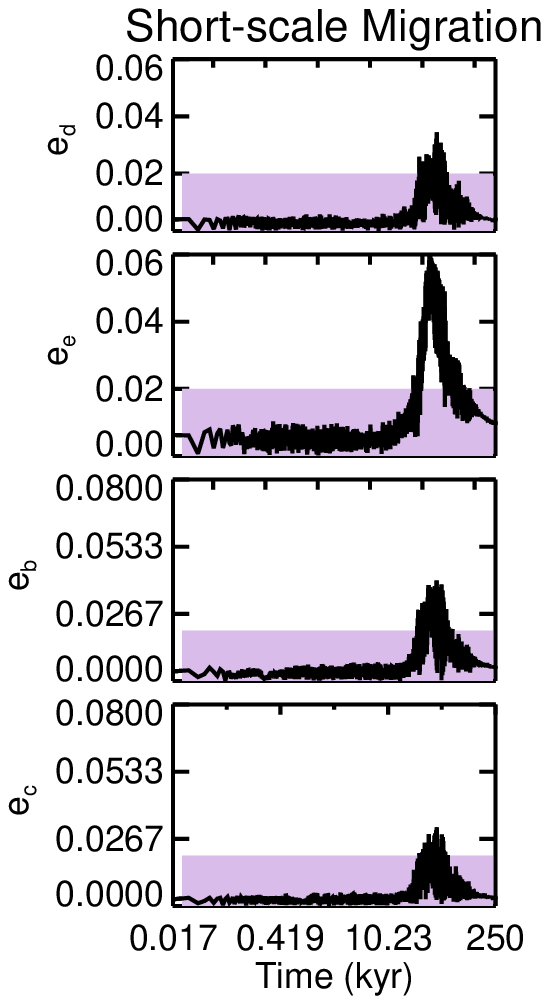}
    \includegraphics[width=0.3\textwidth]{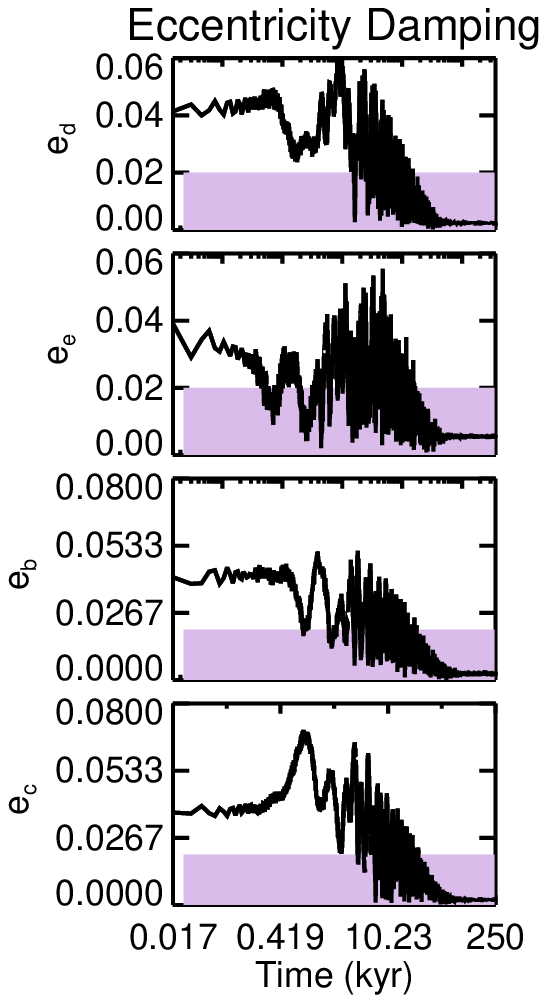}
    \caption{Each dynamical history can produce the eccentricities of Kepler-80 that are smaller than the upper limit imposed by \citet{MacDonald2016}. We plot the estimated ranges for the eccentricities in purple.}
    \label{fig:k80_2}
\end{figure*}

\begin{figure*}[h]
    \centering
    \includegraphics[width=0.3\textwidth]{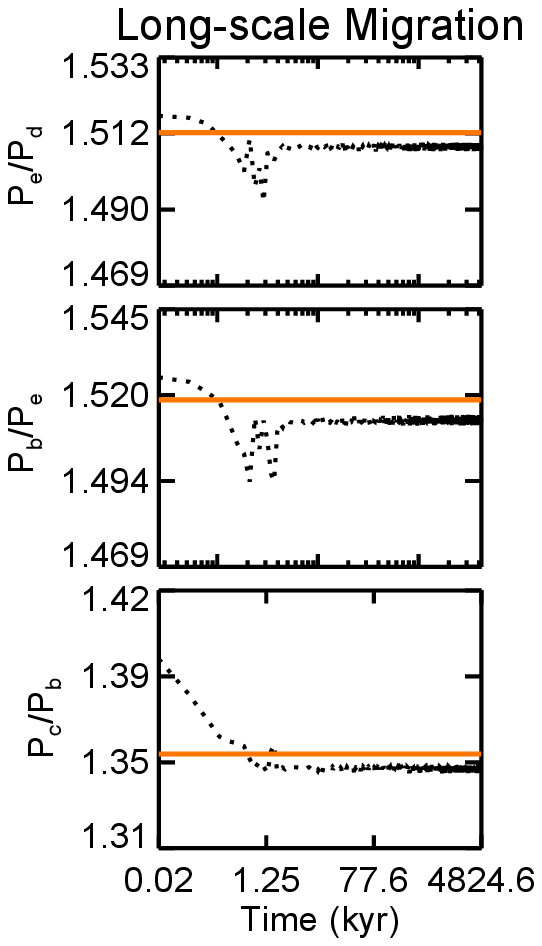}
    \includegraphics[width=0.3\textwidth]{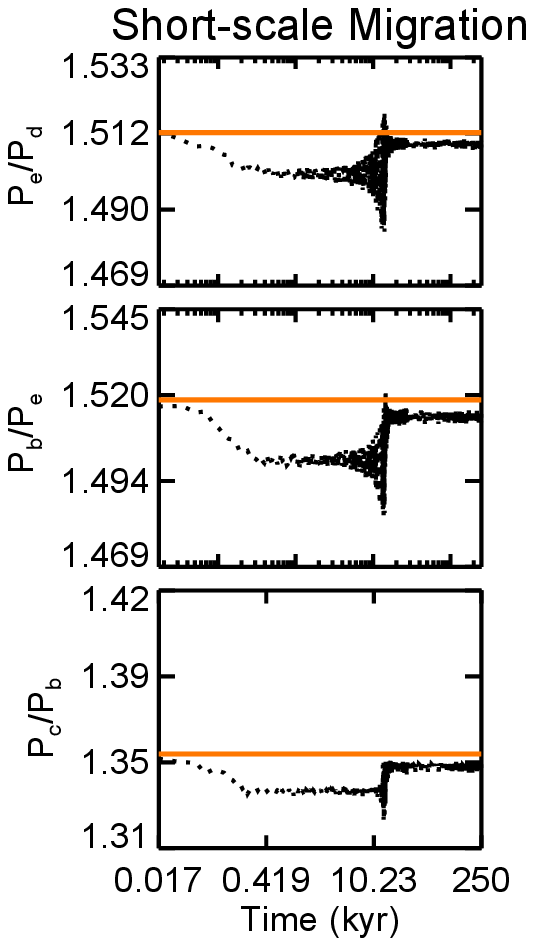}
    \includegraphics[width=0.3\textwidth]{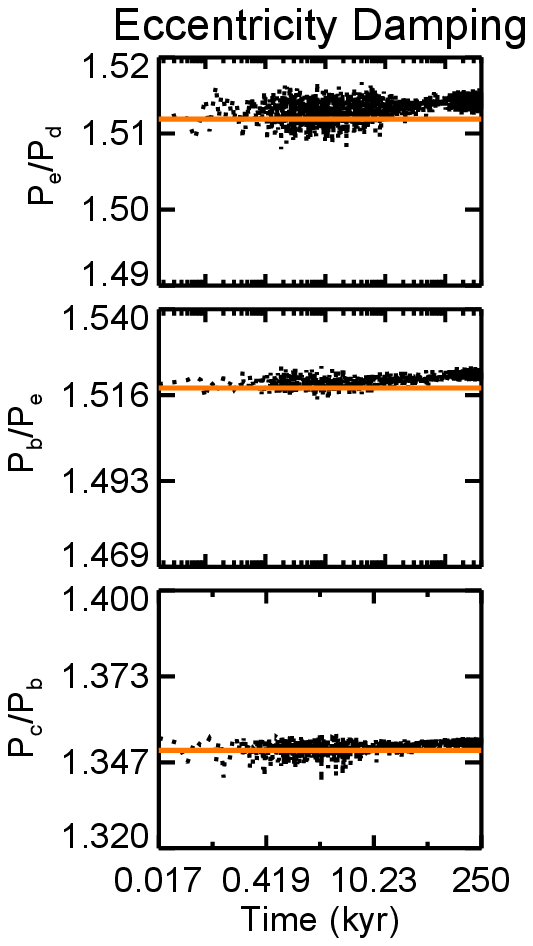}
    \caption{Each dynamical history can produce period ratios between adjacent planets that are similar to those observed in Kepler-80 \citep{MacDonald2016}. We plot the observed values overtop in orange.}
    \label{fig:k80_3}
\end{figure*}

We reproduce the orbital resonances in Kepler-80 via long-scale migration, short-scale migration, and eccentricity damping. We show examples of each of the dynamical histories in Figures \ref{fig:k80}, and compare the eccentricities and period ratios from the simulations to  the observational constraints in Figures \ref{fig:k80_2} and \ref{fig:k80_3}. Since the eccentricities in \citet{MacDonald2016} were forced to be below 0.03, we plot 0.03 as an upper limit for all eccentricities. 

All three dynamical histories were able to reproduce the period ratios of adjacent planets observed in the system, but the long-scale migration simulations typically had period ratios that were smaller than those observed, and the eccentricity damping simulations usually had period ratios that were larger than those observed (e.g., Figure \ref{fig:k80_2}). The short-scale migration simulations resulted in a variety of period ratios, some that were larger than the observed values and others that were smaller.

\begin{deluxetable}{rcccccc}
\tablecaption{Libration Angles for Kepler-80\label{tab:K80res}}
\tablecolumns{7}
\tablewidth{0.43\textwidth}
\tabletypesize{\footnotesize}
\tablehead{\colhead{} & \colhead{$\phi$} & \colhead{$\phi_c$} & \colhead{$\sigma_{\phi_c}$} & \colhead{$\phi_a$} & \colhead{$\sigma_{\phi_a}$} & \colhead{Num.} }   

\startdata
\textbf{lm}  & $\phi_1$  & -152.85 & 1.55 & 6.68 & 7.85 & 21  \\
             & $\phi_1$  & -158.01 & 1.62 & 1.82 & 2.89 & 36  \\
             & $\phi_2$  & -71.63 & 0.95 & 1.11 & 1.28 & 59  \\
             
\textbf{sm}  & $\phi_1$  & -158.99 & 4.97 & 8.50 & 18.96 & 188  \\
             & $\phi_2$  & -71.22 & 1.38 & 3.95 & 3.65 & 188  \\
             
\textbf{ecc} & $\phi_1$  & -4.88 & 5.66 & 93.35 & 6.50 & 13  \\
             & $\phi_1$  & -160.89 & 20.58 & 91.68 & 9.49 & 6  \\
             & $\phi_2$  & -5.03 & 4.03 & 77.13 & 6.02 & 38  \\
             & $\phi_2$  & -73.46 &  4.73 & 22.81 & 8.66 & 47 
\enddata
\tablecomments{Resulting libration centers ($\phi_c$) and amplitudes ($\phi_a$) from our simulations, as well as the number of angles at each center. Here, $\phi_1=2\lambda_e-3\lambda_b+1\lambda_c$ and $\phi_2=3\lambda_d-5\lambda_e+2\lambda_b$ denote the three-body angles between the innermost three planets and the outermost three planets, respectively. The nominal value for each angle and parameter is taken from the mean of the centers and amplitudes of the surviving simulations in resonance, and the uncertainties are taken from the standard deviation. `lm', `sm' and `ecc' stand for long-scale migration, short-scale migration, and eccentricity damping, respectively. Angles are repeated when there are multiple libration centers. All parameters are in degrees and wrapped between [-180,180]. For $\phi_1$ from the eccentricity damping simulations, there was no common center between the librating angles. Because of this, we do not include statistics from the simulations, but do include the number of simulations in which this angle librated. For the three dynamical histories, the following number of simulations resulted in resonance: 257/300, 80/300, 135/500 for short-scale migration, long-scale migration, and eccentricity damping, respectively. Simulations that were in resonance but did not have their three-body angles librating had all two-body angles librating.}
\end{deluxetable}

For the simulations that survived, we analyze the centers and amplitudes of the librating three-body angles, the results of which we show in Table \ref{tab:K80res}. All of our surviving short-scale migration and long-scale migration simulations that had librating three-body angles librated with a center of $\phi_1 \sim -160^{\circ}$ and amplitude $\sim 2--10^{\circ}$ and $\phi_2 \sim -70^{\circ}$ and amplitude $\sim 2--10^{\circ}$.  The eccentricity damping simulations result in the $\phi_1$ angle librating about this center of $\phi_1 \sim -160^{\circ}$ but also librating around $\sim -5^{\circ}$. The $\phi_2$ angle librated around the center $\phi_2 \sim -70^{\circ}$ that was measured by \citet{MacDonald2016} and reproduced by the other the dynamical histories, but some eccentricity damping solutions result in $\phi_2$ angles that librate around the center $\phi_2 \sim -5^{\circ}$ instead.


All three dynamical histories were able to reproduce the observed resonant chain in Kepler-80, the eccentricities within the range imposed by \citet{MacDonald2016}, and the observed period ratios.


\section{Kepler-223}\label{sec:K223}

Kepler-223 is a slightly-evolved, Sun-like star, hosting a planetary system of four planets. All four planets were confirmed by \citet{Rowe2014}, and orbital properties and masses for all four planets were characterized via TTV interactions by \citet{Mills2016}. The planets have orbital periods ranging from 7.4 to 19.7 days and planetary masses ranging from 4.8--8$M_{\oplus}$.
The four planets are interlocked in a four-body resonance of 3:4:6:8, with the innermost three-body angle librating, as well as the outermost three-body angle and numerous two-body angles. The four-body resonance angle, described by  $\phi_3=3\lambda_b-4\lambda_c-3\lambda_d+4\lambda_e$, was also found to librate.

The three-body angles for this system were measured by \citet{Mills2016} who found the angle $\phi_1=-\lambda_b+2\lambda_c-\lambda_d$ describing the interactions between the innermost three planets (b, c, and d) to be centered at $180^{\circ}$ with an amplitude of $\sim10^{\circ}$ and the angle between the outermost three planets (c, d, and e) 
$\phi_2=\lambda_c-3\lambda_d+2\lambda_e$ to be centered at $65^{\circ}$ with an amplitude of $\sim5^{\circ}$. \citet{Mills2016} also simulated the long-scale migration for the system and reproduced these centers and amplitudes. \citet{Delisle2017} applied an analytic model to Kepler-223 and found six possible equilibrium configurations between the two three-body angles.

\citet{Mills2016} measured the eccentricities of the four planets to be non-zero, with values of $0.078^{+0.015}_{-0.017}$, $0.15^{+0.019}_{-0.051}$, $0.037^{+0.018}_{-0.017}$, and $0.051^{+0.019}_{-0.019}$ for planets b, c, d, and e, respectively.

\citet{Mills2016} argued that long-scale migration was required to lock the planets into their four-body resonance along with the associated three- and two-body resonances. We aim to reproduce this resonant chain by simulating long-scale migration, short-scale migration, and eccentricity damping. For all three suites of simulations, we assume a stellar mass of $M=1.125M_{\odot}$ \citep{Mills2016}. We draw values for planetary masses and inclinations from normal distributions centered on values from \citet{Mills2016}, and set all eccentricities for the short and long-scale migration simulations to zero. For the eccentricity damping simulations, all eccentricities are initialized at the values predicted by \citet{Mills2016}. All other values are chosen as described in Section \ref{sec:simulations}.  

\begin{figure*}[t]
    \centering
    \includegraphics[width=0.33\textwidth]{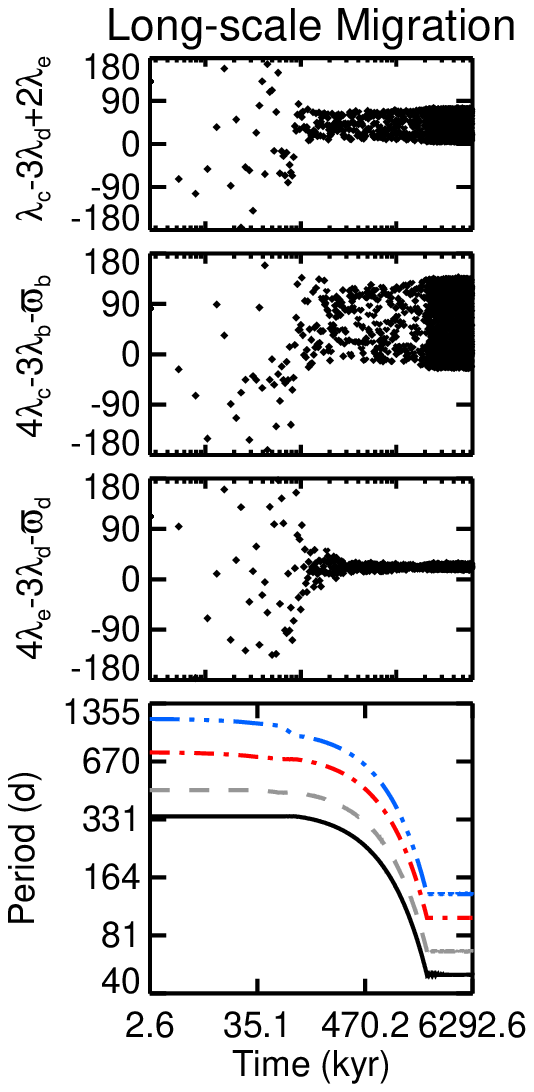}
    \includegraphics[width=0.33\textwidth]{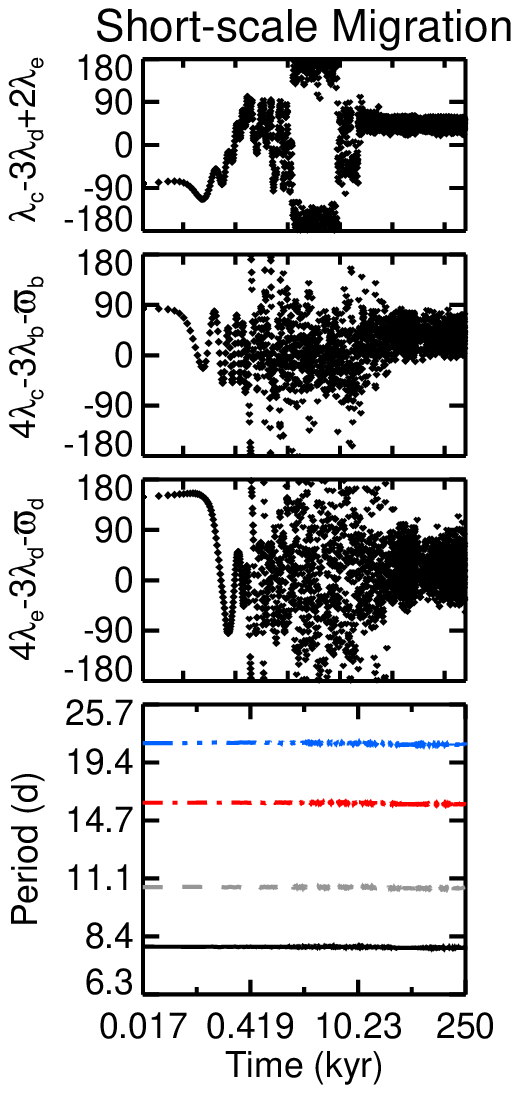}
    \includegraphics[width=0.33\textwidth]{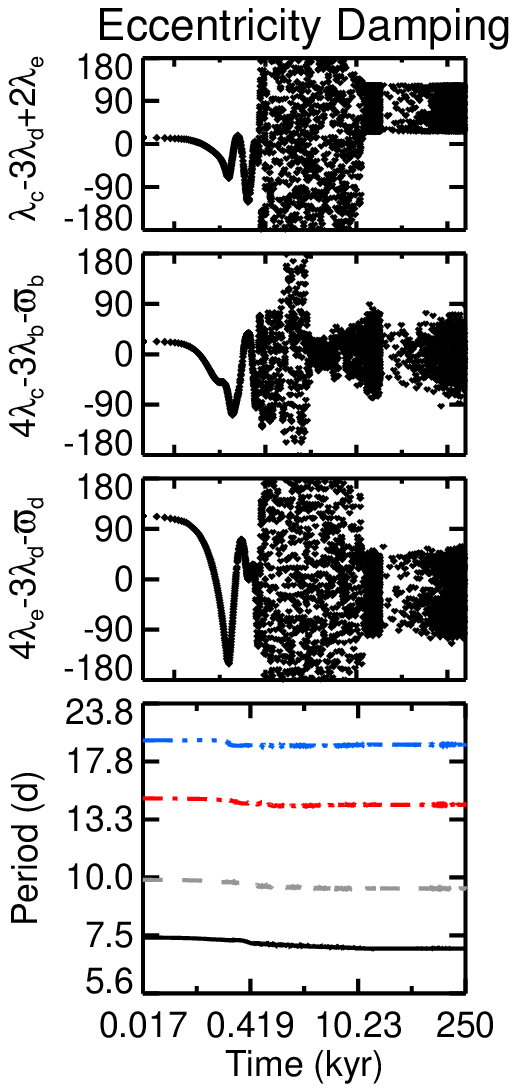}
    \caption{Simulations of each dynamical history can produce the resonant configuration observed in Kepler-223.  Panels from left to right: reproduced via long-scale migration, short-scale migration, and eccentricity damping. Each panel shows the three-body angle $\phi_1$, the two-body angles associated with the three-body angle, and the planets' periods. Note that although only one of the three-body angles is shown here, as well as only two of the various two-body angles, they are all in fact librating. The four-body angle $\phi_3$ also librates. Evolution timescales in days for long-scale migration, short-scale migration, and eccentricity damping: $\tau_a=1.3\times10^8 $, $\tau_e =1.3\times10^6 $, $\tau_a=4.4\times10^7 $, $\tau_e = 2.0\times10^5$, $\tau_e = 2.9\times10^3$.}
    \label{fig:k223}
\end{figure*}

\begin{figure*}[h]
    \centering
    \includegraphics[width=0.3\textwidth]{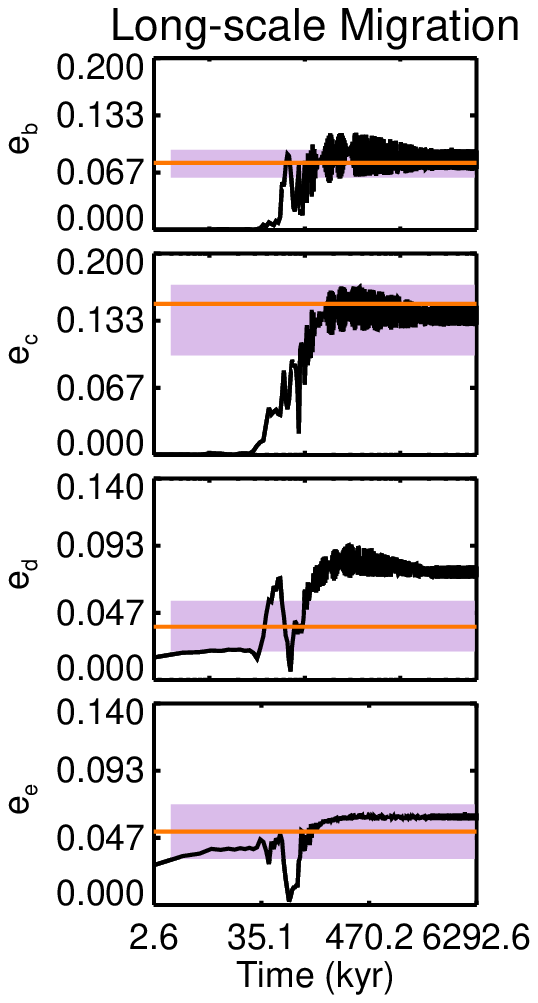}
    \includegraphics[width=0.3\textwidth]{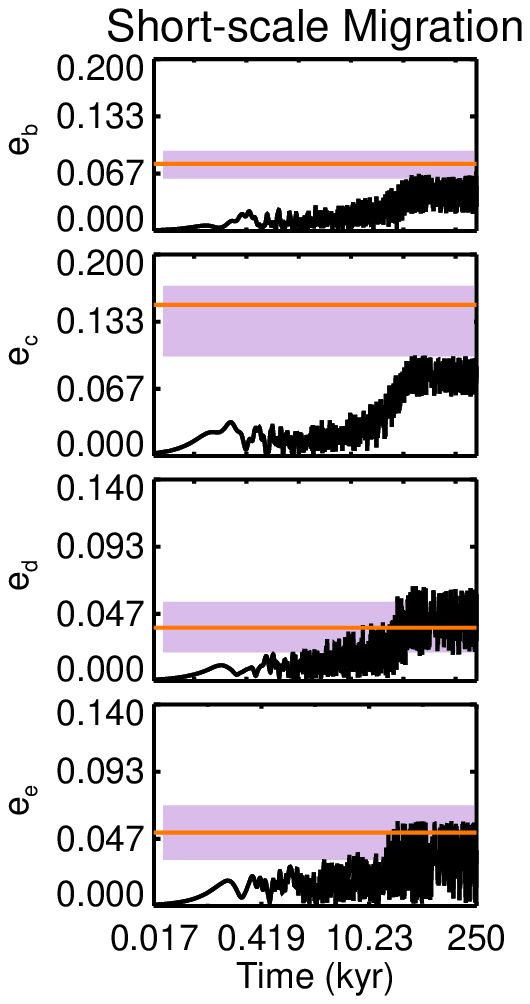}
    \includegraphics[width=0.3\textwidth]{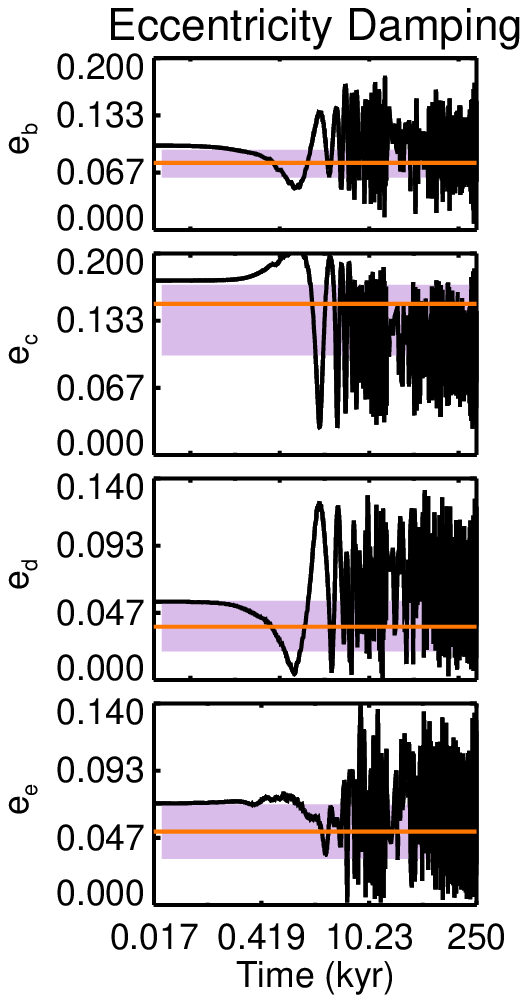}
    \caption{Each dynamical history can produce eccentricities measured in Kepler-223 \citep{Mills2016}. We plot the measured eccentricities overtop in orange and the uncertainties in the measurements in light purple.}
    \label{fig:k223_2}
\end{figure*}

\begin{figure*}[h]
    \centering
    \includegraphics[width=0.3\textwidth]{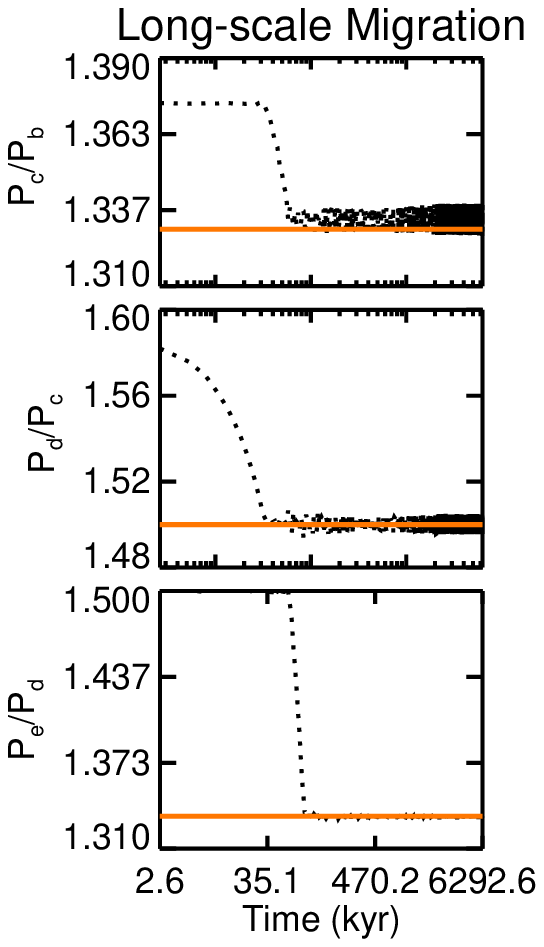}
    \includegraphics[width=0.3\textwidth]{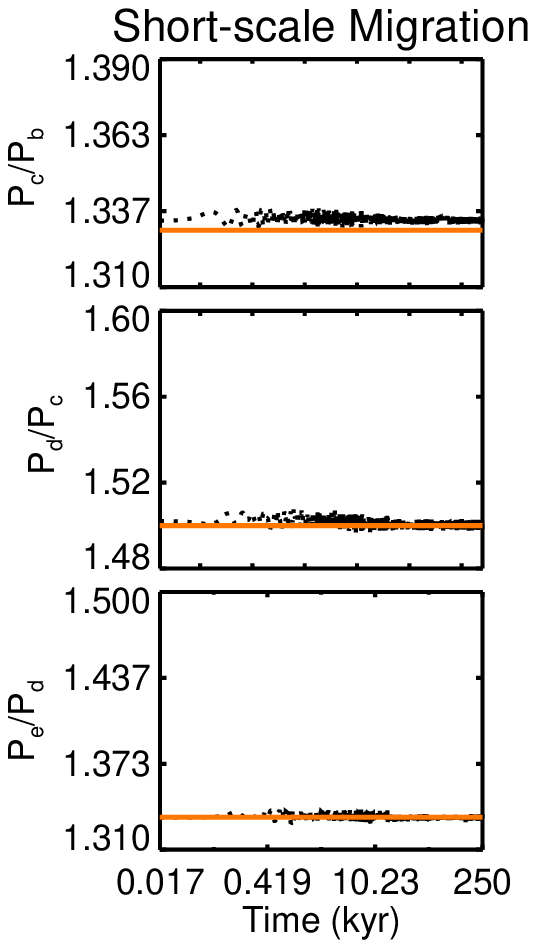}
    \includegraphics[width=0.3\textwidth]{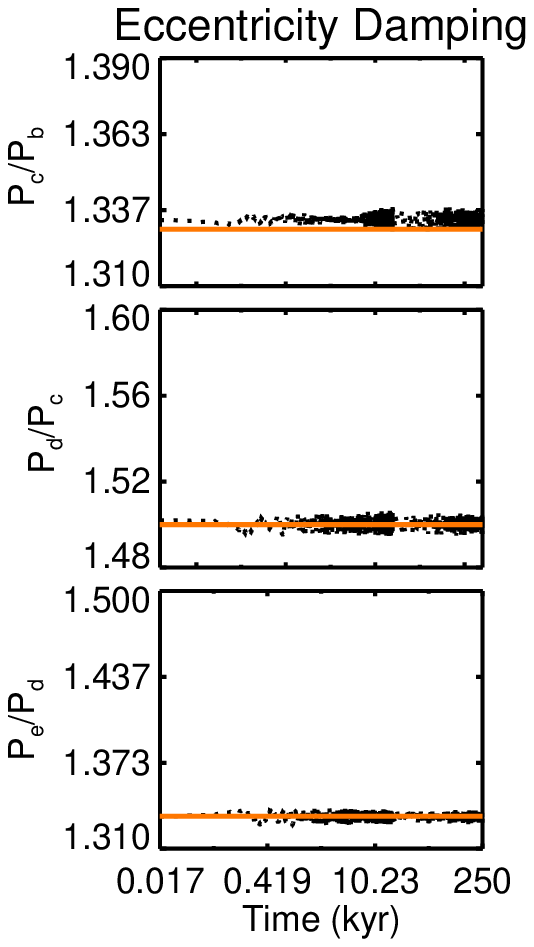}
    \caption{Each dynamical history can produce the period ratios between adjacent planets that are observed in Kepler-223 \citep{Mills2016}. We plot the observed values for the period ratios overtop as orange lines. }
    \label{fig:k223_3}
\end{figure*}

We reproduce the resonant chain observed in Kepler-223 via long-scale migration, short-scale migration, and eccentricity damping only. Examples of resonant angles from simulations of each of the scenarios are shown in Figure \ref{fig:k223}, and we compare the eccentricities and period ratios from the simulations with observational constraints in Figures \ref{fig:k223_2} and \ref{fig:k223_3}. 
For each dynamical history, we found at least one set of initial conditions that recreated non-zero eccentricities for these planets, however, most eccentricity damping simulations resulted in lower eccentricities. Nearly all simulations for all three dynamical histories resulted in a period ratio between planets c and b that was slightly wide of the observed ratio by $\sim$0.1\%.  

For the simulations that survived, we analyze the centers and amplitudes of the three-body angles, the results of which we show in Table \ref{tab:K223res}. Our simulations result in all of the stable configurations previously predicted \citep{Delisle2017}, as well as other locations. Our simulations with short-scale migration resulted in three-body configurations whose $\phi_1$ librates around three different centers (two of which are predicted analytically), and whose $\phi_2$ librates around two different centers (both of which are predicted) with amplitudes ranging between $10^{\circ}$ and $90^{\circ}$. The long-scale migration simulations result in no consistent center for $\phi_2$, and the $\phi_1$ angle librated around $\phi_1\sim45^{\circ}$ (predicted) with amplitudes ranging between $10^{\circ}$ and $45^{\circ}$. Many of the long-scale migration simulations that survived were marked as being out of resonance (in Tables \ref{tab:simnum} and \ref{tab:K223res}), but were actually in a resonance. This is because many of the simulations (76\%) actually locked into a different resonance, where planets d and e locked into a 3:2 resonance (instead of the 4:3) and so the angle $\phi=2\lambda_c-5\lambda_d+3\lambda_e$ librated about 180$^{\circ}$, often with tight amplitudes. Because 3:2 is not the observed resonance, it is not included in the resonance analysis. In the eccentricity damping simulations, the $\phi_1$ angle librated about either $\sim15^{\circ}$ or $\sim180^{\circ}$, and the $\phi_2$ angle librated about $\sim10^{\circ}$ or $\sim175^{\circ}$.
For these three dynamical histories, the four-body angle $\phi_3$ librates in many of the simulations about $\sim0^{\circ}$ with amplitudes ranging from 10--40$^{\circ}$.

\begin{deluxetable}{rcccccc}
\tablecaption{Libration Angles for Kepler-223\label{tab:K223res}}
\tablecolumns{7}
\tablewidth{0.43\textwidth}
\tabletypesize{\footnotesize}
\tablehead{\colhead{} & \colhead{$\phi$} & \colhead{$\phi_c$} & \colhead{$\sigma_{\phi_c}$} & \colhead{$\phi_a$} & \colhead{$\sigma_{\phi_a}$} & \colhead{Num.}}   

\startdata
\textbf{lm}  & $\phi_1$  & 44.06 & 12.60 & 25.91 & 18.25 & 26  \\
             & $\phi_2$  & \nodata & \nodata & \nodata & \nodata &  5 \\
             
\textbf{sm}  & $\phi_1$  & 65.74   & 12.80   & 10.94 & 5.32 & 31  \\
             & $\phi_1$  & 172.56  & 13.71    & 76.63 & 33.84 & 44  \\
             & $\phi_2$  & 7.72    & 5.12     & 76.33 & 8.94 & 36  \\
             & $\phi_2$  & 61.16   & 4.46    & 16.77 & 9.23 & 78  \\
             & $\phi_2$  & 81.64   & 7.37     & 14.64 & 8.50 & 99  \\
             
\textbf{ecc} & $\phi_1$  & 15.35   & 13.55    & 70.78 & 25.42 & 19  \\
             & $\phi_1$  & 180.56  & 6.23     & 93.45 & 4.425 & 9  \\
             & $\phi_2$  & 10.24   & 7.29     & 83.66 & 6.82 & 33  \\
             & $\phi_2$  & 176.01  & 9.19     & 90.35 & 4.60 & 9 
\enddata
\tablecomments{Resulting libration centers ($\phi_c$) and amplitudes ($\phi_a$) from our simulations, as well as the number of angles at each center. Here, $\phi_1=-\lambda_b+2\lambda_c-\lambda_d$ and $\phi_2=\lambda_c-3\lambda_d+2\lambda_e$ denote the three-body angles between the inner three and outer three planets. `lm', `sm' and `ecc' stand for long-scale migration, short-scale migration, and eccentricity damping, respectively.  Angles are repeated when there are multiple libration centers. All parameters are in degrees and wrapped between [0,360]. See Table \ref{tab:K80res} for a description of variables. For $\phi_2$ from the long-scale migration simulations, there was no common center between the librating angles. Because of this, we do not include statistics from the simulations, but do include the number of simulations in which this angle librated. For the three dynamical histories, the following number of simulations resulted in resonance: 205/300, 28/300, 19/300 for short-scale migration, long-scale migration, and eccentricity damping, respectively. Simulations that were in resonance but did not have their three-body angles librating had all two-body angles librating. Many simulations were not in the observed resonance as planets d and c locked into the 3:2 resonance instead of the expected 4:3 and were not included in this table.}
\end{deluxetable}

All three dynamical histories were able to reproduce the observed resonant chain in Kepler-223, the eccentricities within the range measured by \citet{Mills2016}, and the observed period ratios.


\section{Kepler-60}\label{sec:K60}

Kepler-60 is a Sun-like star, hosting a planetary system of three transiting planets with orbital periods ranging from 7.1 to 11.9 days and planetary masses ranging from 4.1--4.5$M_{\oplus}$. The three planets were confirmed by \citet{Steffen2013} via anti-correlated TTVs, who found them to likely be in a three-body resonance, or in a resonant chain consisting of a 5:4 and a 4:3 two-body resonance. 
However, using solutions from TTV fitting, \citet{Jontof-Hutter2016} found stable, non-resonant solutions in addition to resonant solutions.
Prior to this study, Kepler-60 had not been simulated using any of the dynamical histories explored herein in an attempt to lock the planets into the resonant chain.

\citet{Papaloizou2015} performed N-body simulations on Kepler-60 to study the evolution of the system with the planets tidally damped from the star. In the simulations where the planets locked into resonance, all $\phi_1=\lambda_b-2\lambda_c+\lambda_d$ resonant angles librated around $0^{\circ}$ or $180^{\circ}$, with an amplitude of $\sim50^{\circ}$. Both \citet{Gozdziewski2016} and \citet{Jontof-Hutter2016} performed TTV fitting on the system, and integrated their solutions forward to assess stability and whether the system was in resonance. \citet{Gozdziewski2016} analyzed the system and its TTVs, but found a different libration center of $45^{\circ}$ with an amplitude of about $10^{\circ}$. \citet{Jontof-Hutter2016} examined the three-body angle from their TTV fits and found that only some of the solutions are in resonance. 20\% of their solutions led to a non-librating three-body angle. Of the solutions that were locked into resonance, 98\% of them had the three-body angle librating around $45^{\circ}$ while the remaining solutions had the three-body resonant angle librating about $135^{\circ}$. \citet{Jontof-Hutter2016} also measured the eccentricities of the three planets in the system through their TTV analysis: $e_bcos\omega_b=0.023^{+0.067}_{-0.069}$, $e_bsin\omega_b=0.008 ^{+0.060}_{-0.059}$,  $e_bcos\omega_c= -0.003^{+0.062}_{-0.063}$,$e_bsin\omega_ca=0.034 ^{+0.054}_{-0.053}$, $e_dcos\omega_d= 0.021^{+0.052}_{-0.053}$, and $e_dsin\omega_d= 0.002^{+0.047}_{-0.046}$.

We simulate the planets' orbital evolution and apply long-scale migration, short-scale migration, and eccentricity damping to lock the planets into their resonant chain. For all three suites of simulations, we assume $M=1.105M_{\odot}$ \citep{Rowe2015}. We draw values for planetary masses and inclinations from normal distributions centered on values from \citet{Jontof-Hutter2016}, and all initial eccentricities for short and long-scale migration are zero. For the eccentricity damping simulations, the eccentricities were initialized at 0.05 for all planets. All other values are chosen as described in Section \ref{sec:simulations}. 

\begin{figure*}[t]
    \centering
    \includegraphics[width=0.33\textwidth]{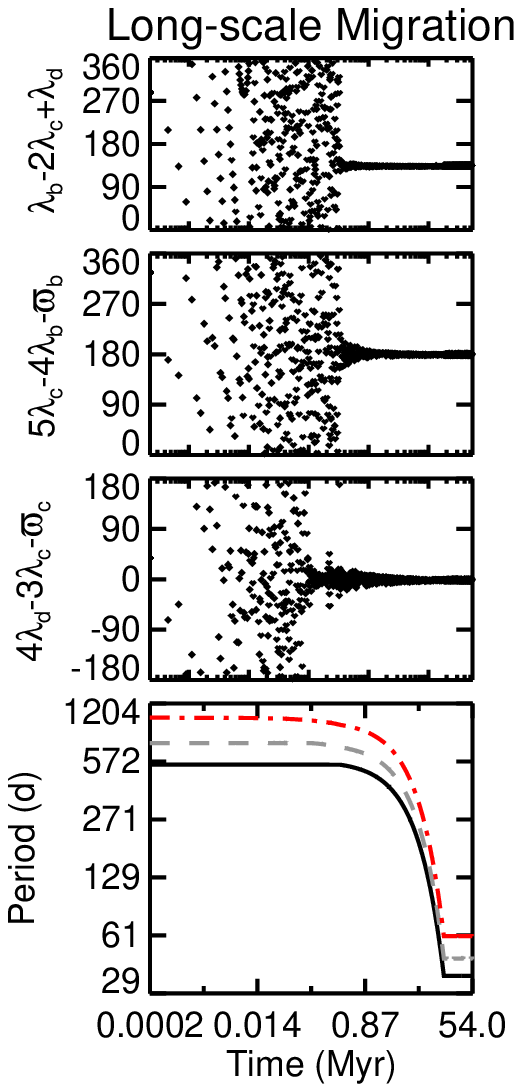}
    \includegraphics[width=0.33\textwidth]{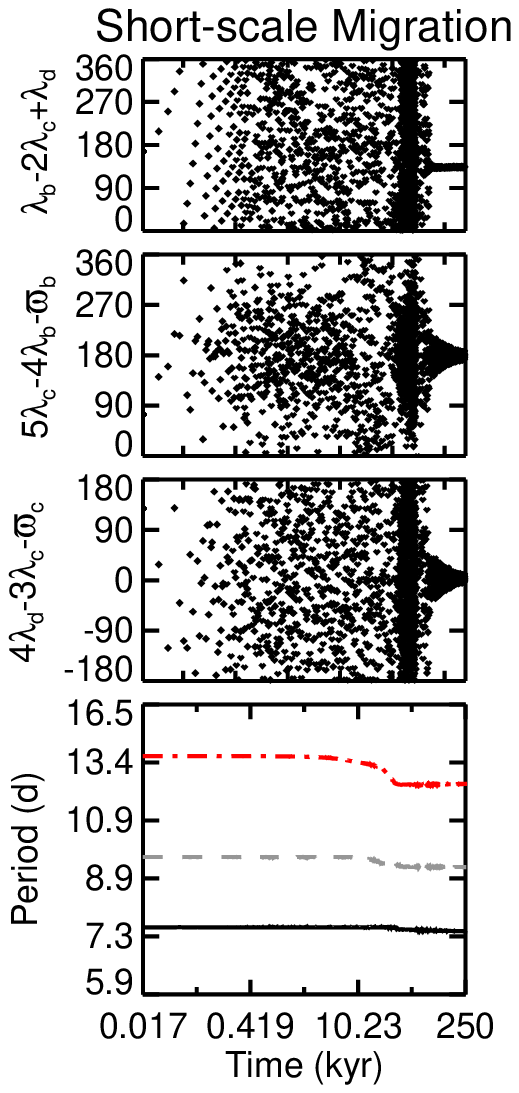}
    \includegraphics[width=0.33\textwidth]{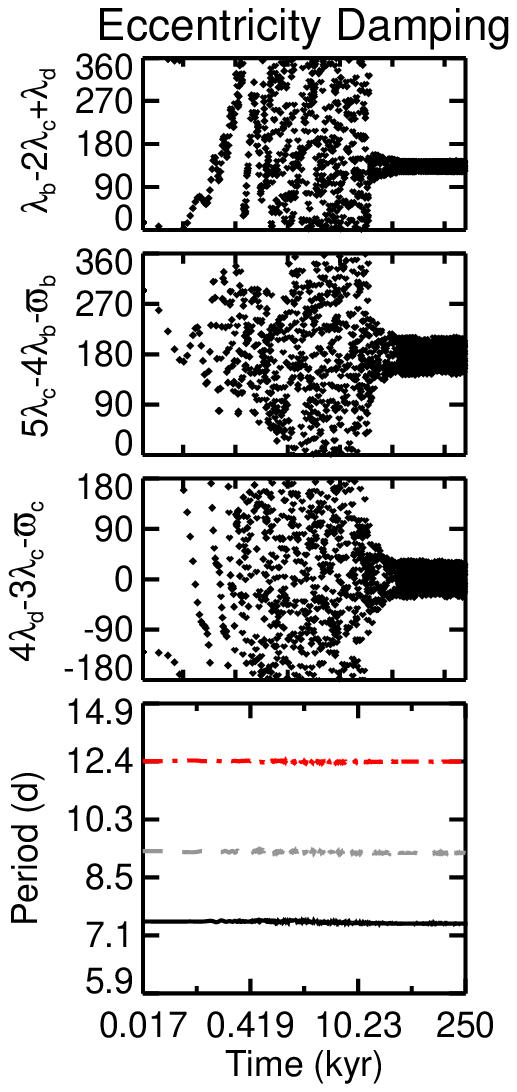}
    \caption{Simulations of each dynamical history can produce the resonant configuration observed in Kepler-60. Panels from left to right: reproduced via long-scale migration, short-scale migration, and eccentricity damping. Each panel shows the three-body angle $\phi_1$, the two-body angles associated with the three-body angle, and the planets' periods. Note that although only two of the various two-body angles are shown, they are all in fact librating. Evolution timescales in days for long-scale migration, short-scale migration, and eccentricity damping: $\tau_a=1.6\times10^9 $, $\tau_e =1.7\times10^5 $, $\tau_a=4.5\times10^6 $, $\tau_e = 3.3\times10^4$, $\tau_e = 4.2\times10^5$.}
    \label{fig:k60}
\end{figure*}

\begin{figure*}[h]
    \centering
    \includegraphics[width=0.3\textwidth]{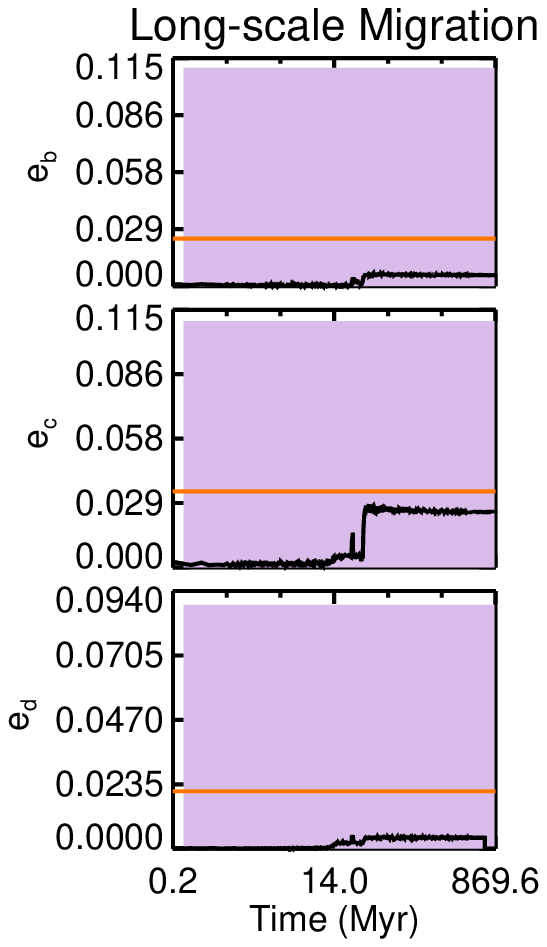}
    \includegraphics[width=0.3\textwidth]{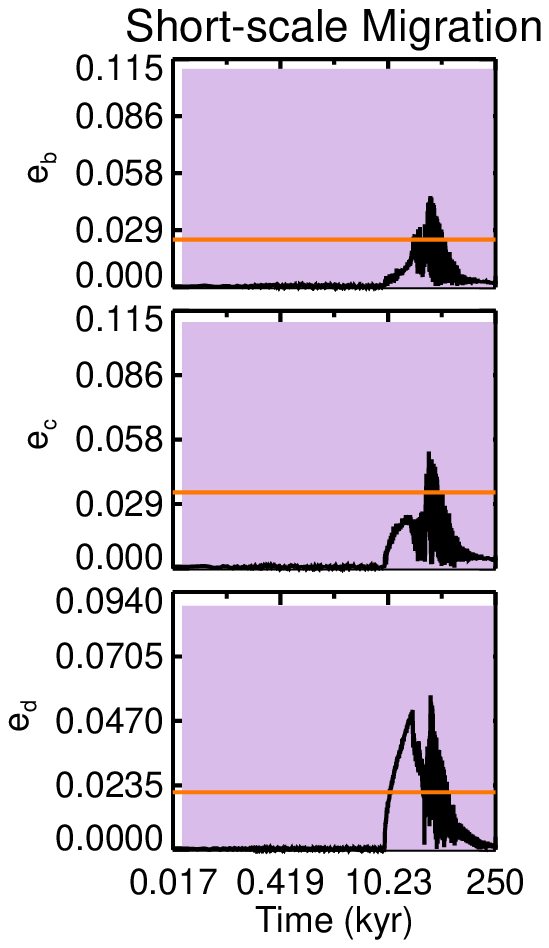}
    \includegraphics[width=0.3\textwidth]{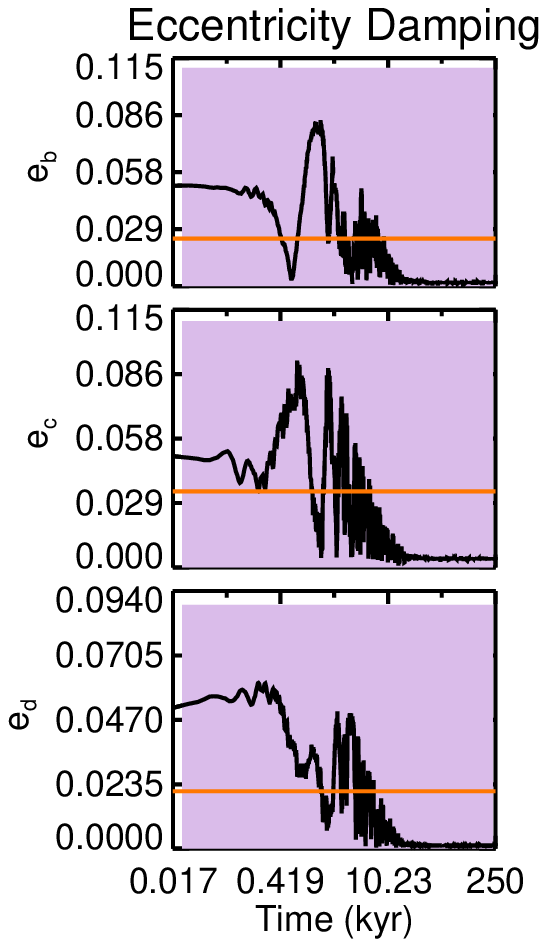}
    \caption{Each dynamical history can produce the eccentricities measured in Kepler-60 \citet{Jontof-Hutter2016}. We plot the measured eccentricities overtop in orange  and the uncertainties for the eccentricities in light purple.}
    \label{fig:k60_2}
\end{figure*}

\begin{figure*}[h]
    \centering
    \includegraphics[width=0.3\textwidth]{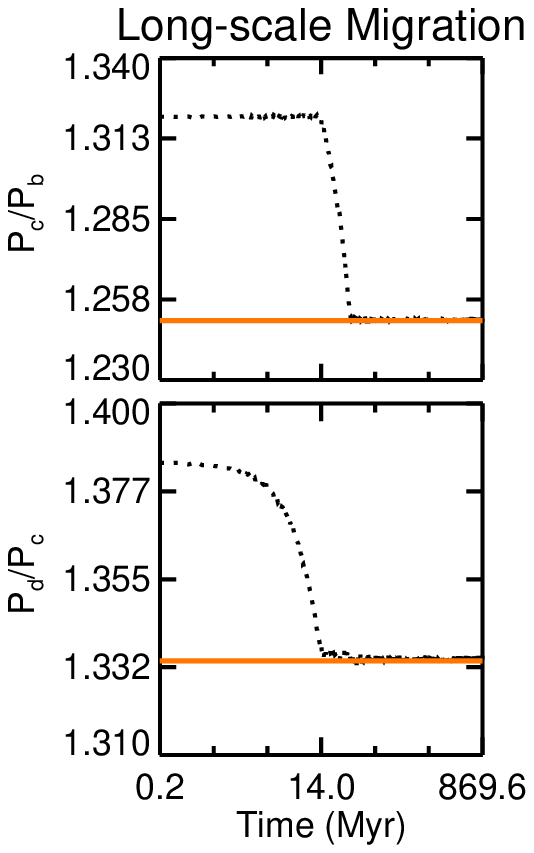}
    \includegraphics[width=0.3\textwidth]{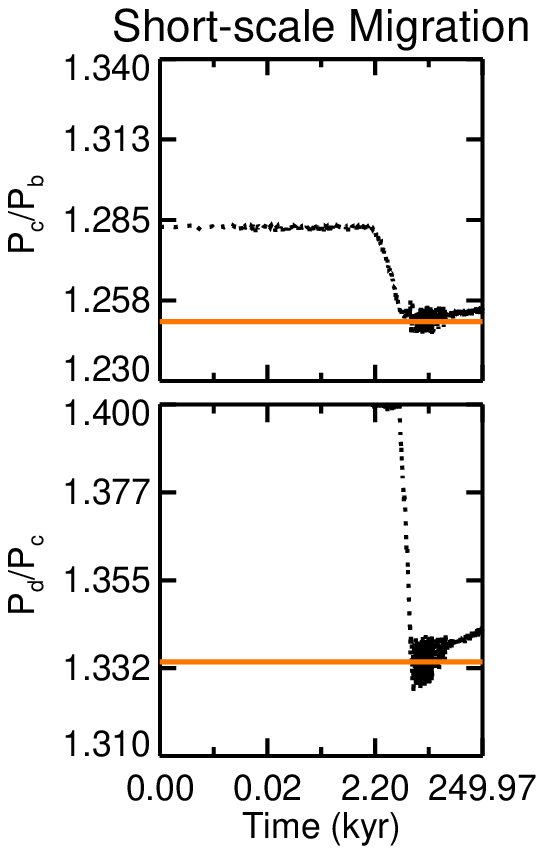}
    \includegraphics[width=0.3\textwidth]{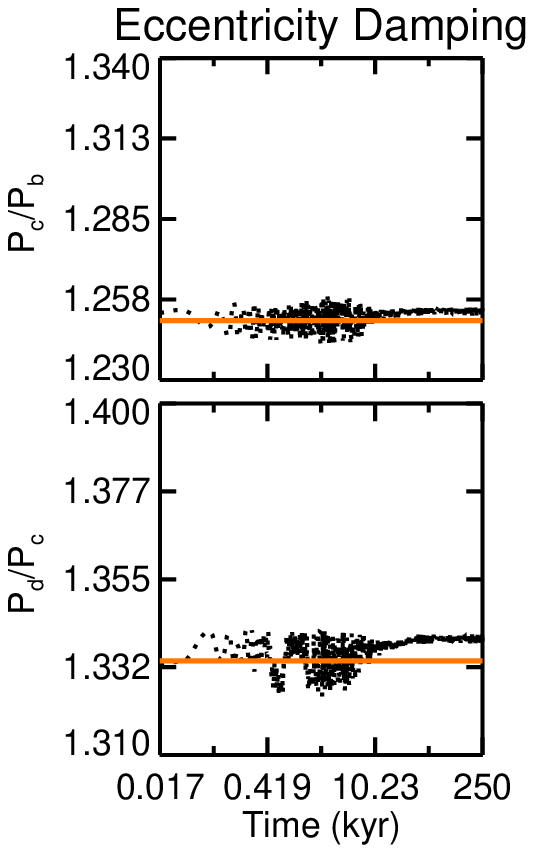}
    \caption{Each dynamical history can produce period ratios that are similar to those observed in Kepler-60 \citep{Jontof-Hutter2016}. We overplot the observed values in orange. }
    \label{fig:k60_3}
\end{figure*}

We reproduce the resonant chain in Kepler-60 via long-scale migration, short-scale migration, and eccentricity damping. Examples of resonant angles from simulations of each of the scenarios are shown in Figure \ref{fig:k60}, and we compare the eccentricities and period ratios from the simulations with observational constraints in Figures \ref{fig:k60_2} and \ref{fig:k60_3}. We plot the nominal eccentricity values as well as their uncertainties in Figure \ref{fig:k60_2}. Nearly all of our simulations from all three dynamical histories resulted in near-circular orbits, below the nominal values but still within uncertainties, and period ratios of adjacent planets that were similar to the observed values. 

For the simulations that survived, we analyze the centers and amplitudes of the librating three-body angle $\phi_1 = \lambda_b-2\lambda_c+\lambda_d$, the results of which we show in Table \ref{tab:K60res}. Each simulation librated about $\sim40^{\circ}$, $\sim134^{\circ}$, $\sim225^{\circ}$, or $\sim316^{\circ}$ with amplitudes between $<1 - 30^{\circ}$. 

\begin{deluxetable}{rcccccc}
\tablecaption{Libration Angles for Kepler-60\label{tab:K60res}}
\tablecolumns{7}
\tablewidth{0.43\textwidth}
\tabletypesize{\footnotesize}
\tablehead{\colhead{} & \colhead{$\phi$} & \colhead{$\phi_c$} & \colhead{$\sigma_{\phi_c}$} & \colhead{$\phi_a$} & \colhead{$\sigma_{\phi_a}$} & \colhead{Num.}}   
\startdata
\textbf{lm}  & $\phi_1$  & 44.68 & 0.25 & 0.84 & 0.98 & 14  \\
             & $\phi_1$  & 134.84 & 0.23 & 0.84 & 1.33 & 33  \\
             & $\phi_1$  & 225.18 & 0.24 & 0.95 & 1.20 & 9  \\
             & $\phi_1$  & 315.59 & 0.28 & 0.47 & 0.36 & 5 \\
             
\textbf{sm}  & $\phi_1$  & 41.13 & 6.10 & 7.04 & 15.93 & 27  \\
             & $\phi_1$  & 134.13 & 3.24 & 3.00 & 8.89 & 29  \\
             & $\phi_1$  & 226.41 & 0.97 & 3.80 & 9.61 & 10  \\
             & $\phi_1$  & 317.30 & 2.39 & 1.41 & 1.25 & 6  \\
             
\textbf{ecc} & $\phi_1$  & 41.22 & 6.14 & 11.38 & 21.24 & 114  \\
             & $\phi_1$  & 134.21 & 1.76 & 6.84 & 10.25 & 86  \\
             & $\phi_1$  & 225.03 & 3.61 & 8.10 & 15.37 & 35 \\
             & $\phi_1$  & 317.48 & 1.21 & 4.65 & 4.82 & 24 
\enddata
\tablecomments{Resulting libration centers ($\phi_c$) and amplitudes ($\phi_a$) from our simulations, as well as the number of angles at each center. Here, $\phi_1=\lambda_b-2\lambda_c+\lambda_d$ donotes the three-body angle between the three planets. `lm', `sm' and `ecc' stand for long-scale migration, short-scale migration, and eccentricity damping, respectively. Angles are repeated when there are multiple libration centers. All parameters are in degrees and wrapped between [0,360]. See Table \ref{tab:K80res} for a description of variables. For the three dynamical histories, the following number of simulations resulted in resonance: 85/300, 72/300, 146/300 for short-scale migration, long-scale migration, and eccentricity damping, respectively. Simulations that were in resonance but did not have their three-body angle librating had all two-body angles librating.}
\end{deluxetable}

All three dynamical histories were able to reproduce the observed resonant chain in Kepler-60, the eccentricities within the range measured by \citet{Jontof-Hutter2016}, and the observed period ratios.


\section{TRAPPIST-1}\label{sec:trap1}

TRAPPIST-1 is a late M-dwarf hosting a system of seven planets, with planetary masses between 0.09--1.6$M_{\oplus}$ and orbital periods ranging from 1.5--19 days \citep{Wang2017}. The system was  discovered by \citet{Gillon2017}. Several of the planets are in or near the habitable zone and therefore could host liquid water. \citet{Gillon2017} measured the orbital period ratios for the inner six planets to be near the ratios of small integers, suggesting that the planets could be in two-body resonances of 8:5, 5:3, 3:2, 3:2, and 4:3, from b:c to g:f. \citet{Luger2017} followed up the discovery with additional data from K2 and found that the outermost planet (h) was near or in the 3:2 MMR with planet g. In the initial discovery paper and following discussions on the system's formation and evolution, it has generally been assumed that the planets needed to migrate to lock into the resonance (e.g., \citealt{Gillon2017,Luger2017,Tamayo2017,Papaloizou2017,Ormel2017}).

It is difficult to measure any two-body resonances without the eccentricity vector. However, three-body resonances, since they only depend on the mean longitude and not the longitudes of periapse, can be measured. \citet{Luger2017} measured the three-body angles in this system, concluding that all angles are librating. They found the following centers and amplitudes of libration: 
$\phi_1=2\lambda_b-5\lambda_c+3\lambda_d$ librates around 177$^{\circ}$ with an amplitude of 1$^{\circ}$;
$\phi_2=\lambda_c-3\lambda_d+2\lambda_e$ librates around 48.5$^{\circ}$ with an amplitude of 1.5$^{\circ}$;
$\phi_3=2\lambda_d-5\lambda_e+3\lambda_f$ librates around -148$^{\circ}$ with an amplitude of 6$^{\circ}$; and
$\phi_4=\lambda_e-3\lambda_f+2\lambda_g$ librates around -75.5$^{\circ}$ with an amplitude of 3.5$^{\circ}$.

\citet{Ormel2017} modeled the system and simulated long-scale migration. They successfully lock the system into the observed period ratios, but note that many of the pairs got trapped in the 2:1 resonance and required faster migration or large gas densities in the disk to ``skip'' this 2:1 resonance and be captured into a tighter one. \citet{Tamayo2017} argue that this chain could be formed through slow, convergent migration. At the end of their N-body simulations, all two-body angles associated with the period ratios and all three-body angles measured by \citet{Luger2017} librated with small amplitudes. The two-body angles librated about 180$^{\circ}$, while the three-body angles librated about $\sim 155 ^{\circ}$, $\sim 60 ^{\circ}$, $\sim 155 ^{\circ}$, and $\sim 70 ^{\circ}$. \citet{Tamayo2017} mention that the libration centers varied depending on the initial conditions. No study has yet measured the eccentricities of these planets, but \citet{Gillon2017} estimated upper limits.

We simulate the inner six planets and apply long-scale migration, short-scale migration, and only eccentricity damping to lock the planets into the chain of two-body resonances listed above of 3:4:6:9:15:24. For all three suites of simulations, we assume $M=0.08M_{\odot}$ \citep{Gillon2017}. We draw values for planetary masses and inclinations from normal distributions centered on values from \citet{Tamayo2017}, and all initial eccentricities are zero. For the eccentricity damping simulations, all eccentricities are initialized at 0.04. All other values are chosen as described in Section \ref{sec:simulations}. 

\begin{figure*}[t]
    \centering
    \includegraphics[width=0.33\textwidth]{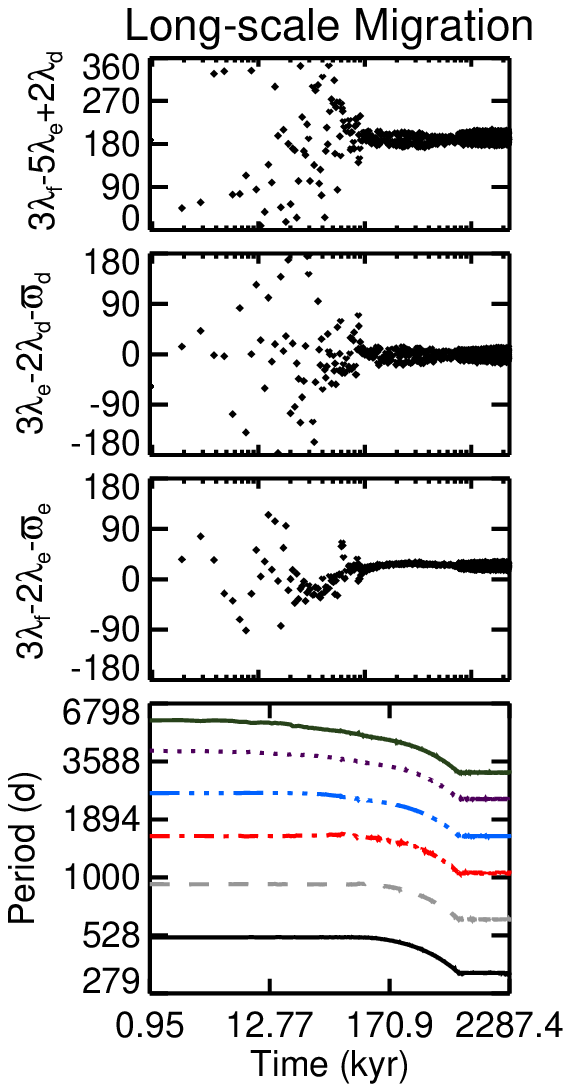}
    \includegraphics[width=0.33\textwidth]{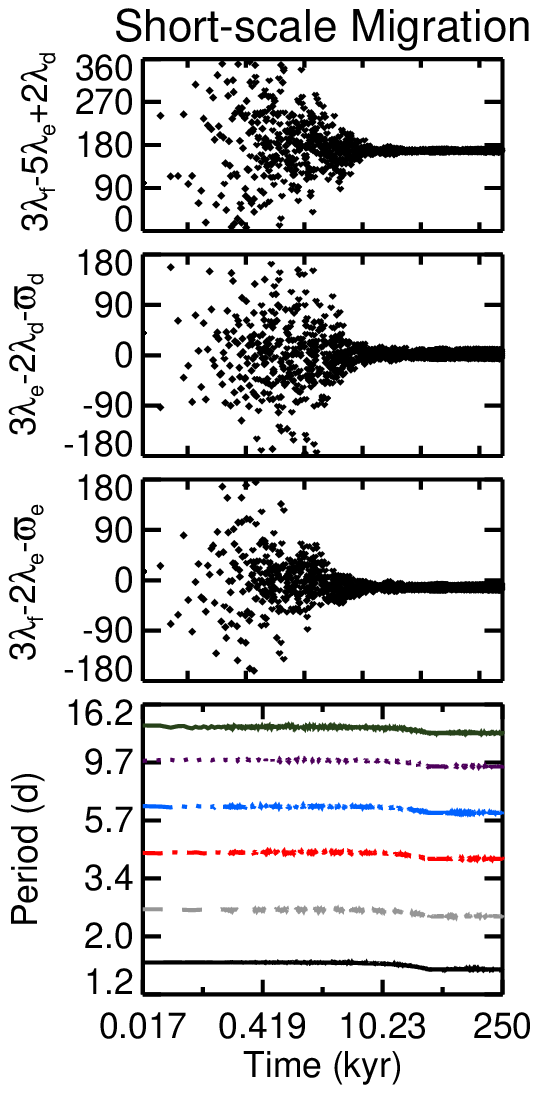}
    \includegraphics[width=0.33\textwidth]{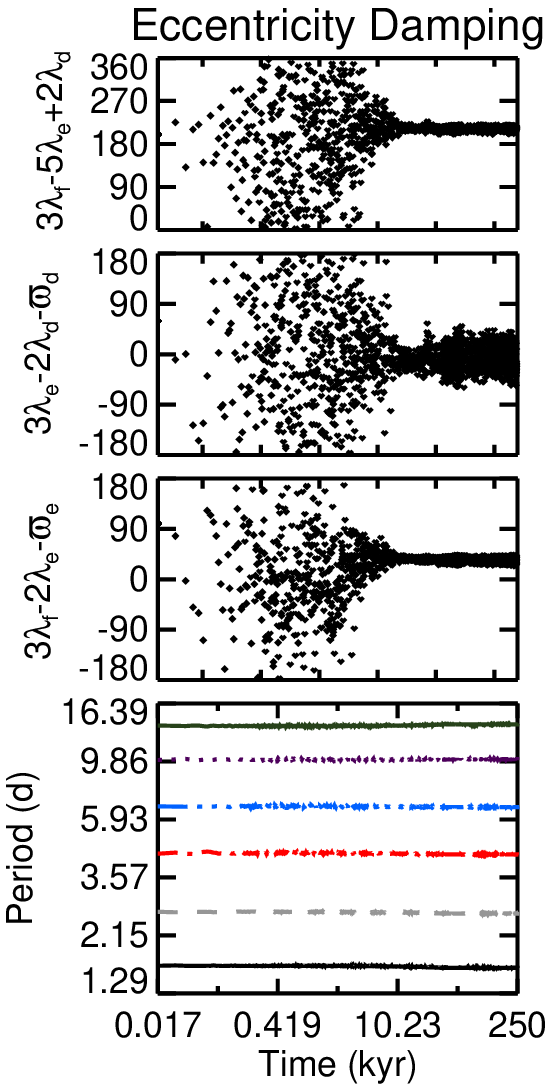}
    \caption{Simulations of each dynamical history can produce the resonant configuration observed in TRAPPIST-1. Panels from left to right: reproduced via long-scale migration, short-scale migration, and eccentricity damping. Each panel shows the three-body angle $\phi_3$, the two-body angles associated with the three-body angle, and the planets' periods. Note that although only one of the three-body angles is shown here, as well as only two of the various two-body angles, they are all librating. Evolution timescales in days for long-scale migration, short-scale migration, and eccentricity damping: $\tau_a=9.3\times10^8 $, $\tau_e =8.2\times10^6 $, $\tau_a=3.5\times10^7 $, $\tau_e = 1.5\times10^4$, $\tau_e = 1.1\times10^4$.}
    \label{fig:trap1}
\end{figure*}

\begin{figure*}[h]
    \centering
    \includegraphics[width=0.3\textwidth]{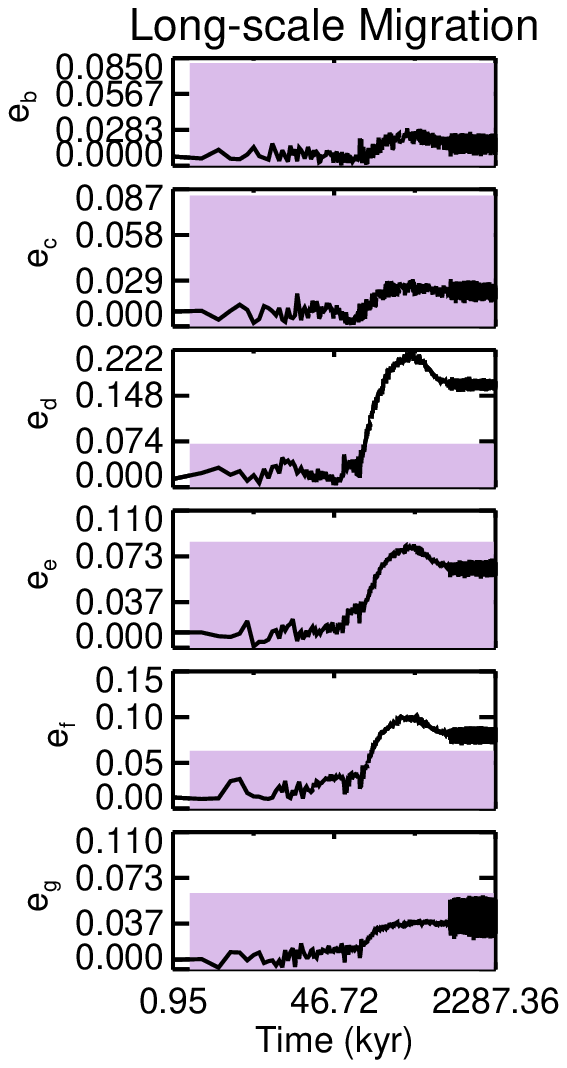}
    \includegraphics[width=0.3\textwidth]{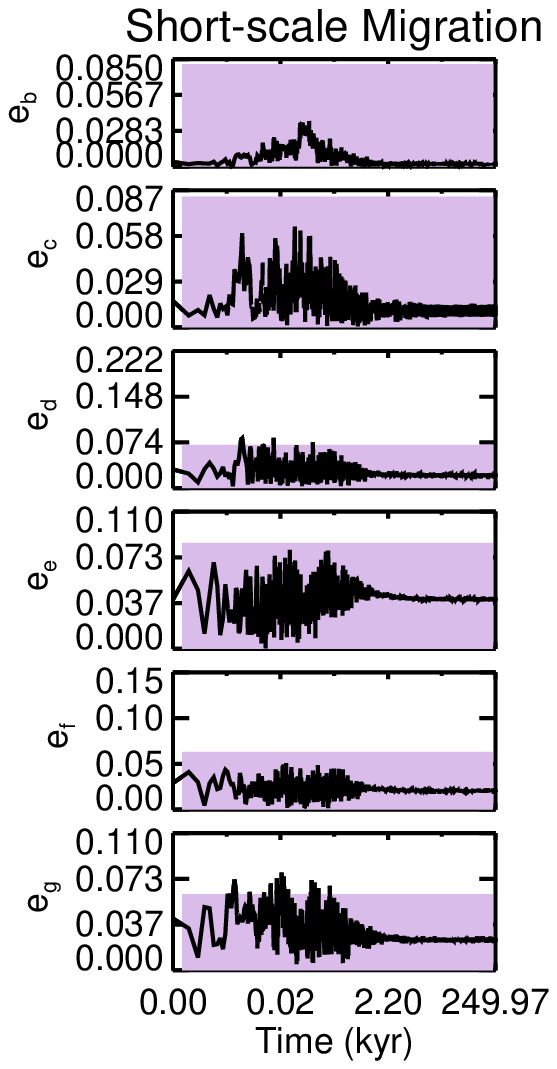}
    \includegraphics[width=0.3\textwidth]{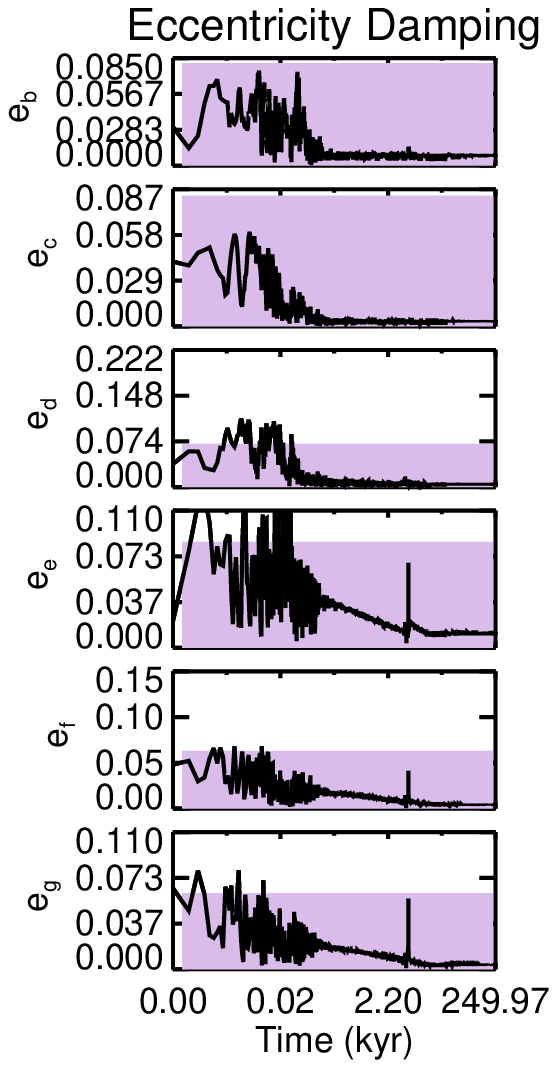}
    \caption{Each dynamical history can produce eccentricities of TRAPPIST-1 smaller than the estimated maximum values \citep{Gillon2017}. We plot the estimated ranges for the eccentricities in purple.}
    \label{fig:trap1_2}
\end{figure*}

\begin{figure*}[h]
    \centering
    \includegraphics[width=0.3\textwidth]{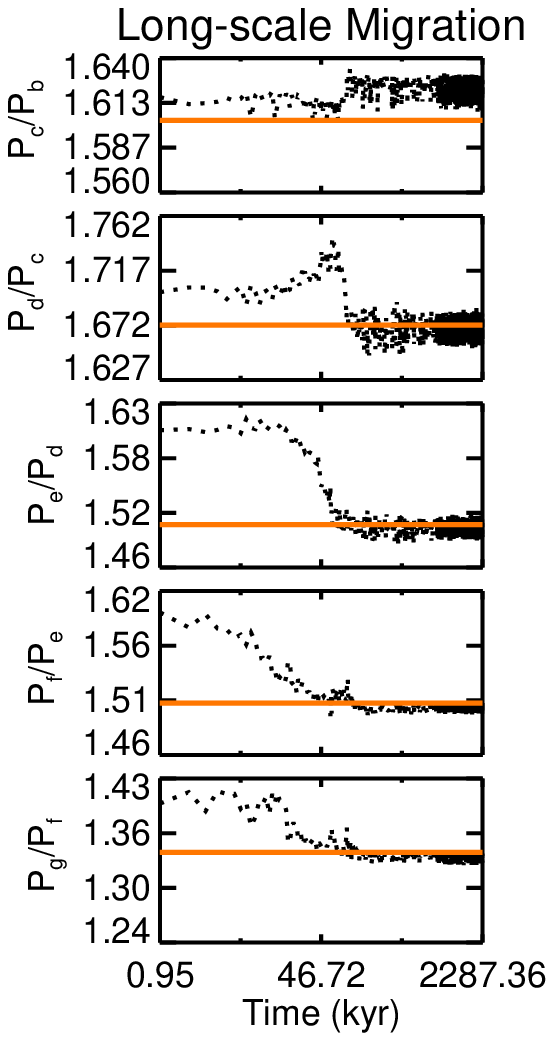}
    \includegraphics[width=0.3\textwidth]{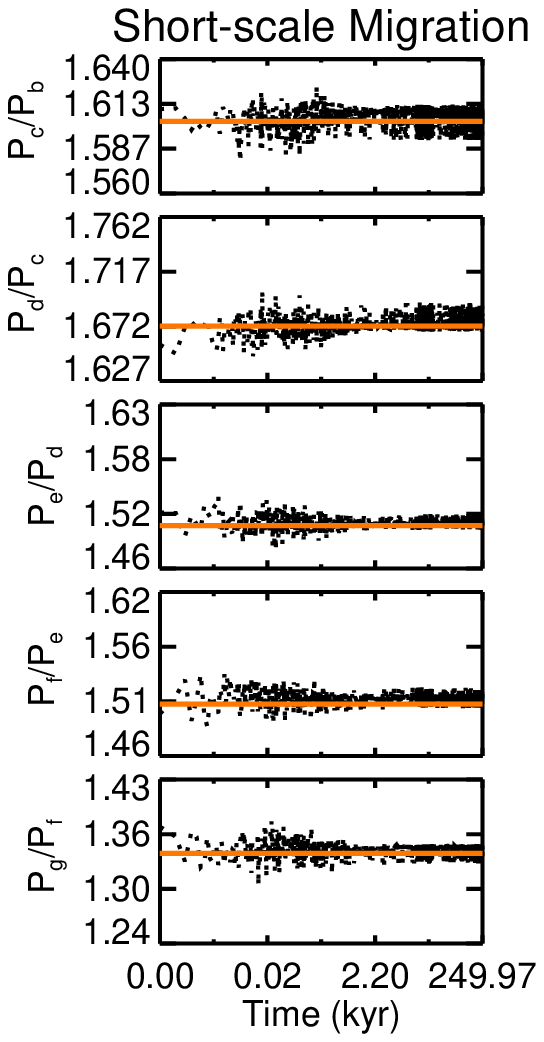}
    \includegraphics[width=0.3\textwidth]{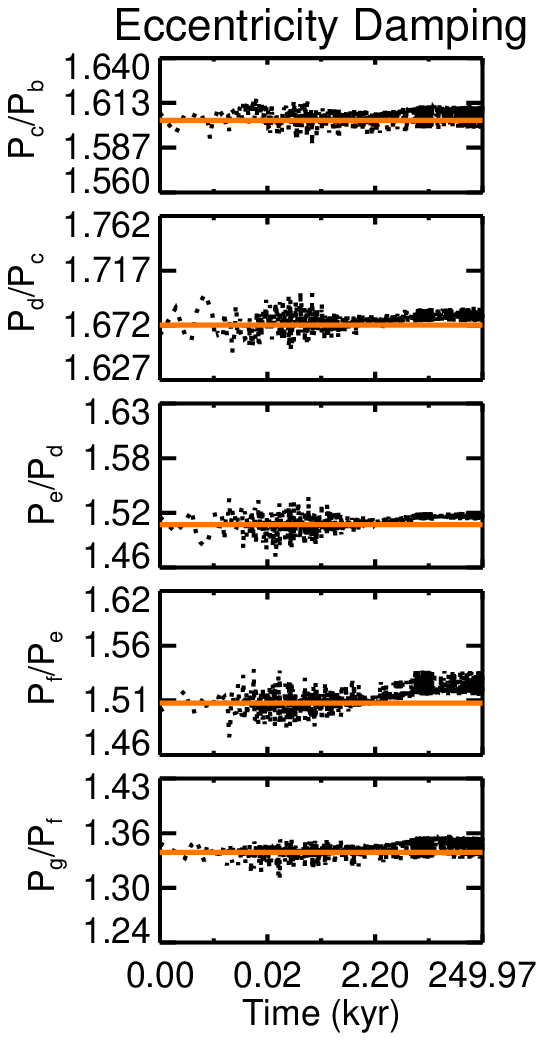}
    \caption{Each dynamical history can produce period ratios of adjacent planets that are similar to those observed in TRAPPIST-1 \citep{Gillon2017}. We plot the observed values overtop in orange.}
    \label{fig:trap1_3}
\end{figure*}

We reproduce TRAPPIST-1 via long-scale migration, short-scale migration, and eccentricity damping. Examples of resonant angles from simulations of each of the scenarios are shown in Figure \ref{fig:trap1}, and we compare the eccentricities and period ratios from the simulations with observational constraints in Figures \ref{fig:trap1_2} and \ref{fig:trap1_3}. We plot the upper limits from \citet{Gillon2017} in Figure \ref{fig:trap1_2} along with the eccentricity evolution of some of our simulations. These limits are rarely breached, except in the case of planets d and f for the long-scale migration simulations for the example shown in Figure \ref{fig:trap1_2}. Other long-scale migrations simulations resulted in the eccentricities of all the planets being within these upper limits, but there were no simulations with all resonant angles librating that had eccentricities smaller than the upper limits\footnote{It is possible that we could find a set of initial conditions that formed the resonant chain as well as matching the eccentricities if we ran more simulations.}. Most simulations from short-scale migration and eccentricity damping result in  eccentricities below the \citet{Gillon2017} upper limits. 

For the simulations that survived, we analyze the centers and amplitudes of the librating three-body angles defined above. We show the resulting average centers and amplitudes as well as their uncertainties of the librating three-body angles from our simulations in Table \ref{tab:trap1res}. Given that only five long-scale migration simulations survived, and only four of them were locked into resonance, we do not have sufficient numbers to analyze the centers and amplitudes for this system for this dynamical history, and they are therefore not included in Table \ref{tab:trap1res}. In fact, angles $\phi_1$, $\phi_2$, and $\phi_4$ only librate in single simulation. $\phi_3$ librates in three simulations, all about a center $\sim160^{\circ}$ with varying amplitudes. For these simulations, the two-body angles between planets b/c rarely lock into their suggested 8:5 resonance. Instead, they lock into 3:2, 9:5, and 7:4. The two-body angles between planets c/d usually lock into their 5:3 resonance, but a few simulations have the 9:5 angle librating instead. At least one simulation, shown in Figure \ref{fig:trap1}, shows the planets locking into the appropriate resonances. For short-scale migration and eccentricity damping, $\phi_1$ librates about $\sim165^{\circ}$, $\phi_2$ librates about $\sim30-45^{\circ}$, $\phi_3$ librates about $\sim140^{\circ}$, and $\phi_4$ librates about $\sim180^{\circ}$. Although there are a variety of similar centers, $\phi_1$, $\phi_2$, and $\phi_3$ probably agree with previous studies, but $\phi_4$ does not agree with the previous results of a center of -75.5$^{\circ}$. Although we were not able to reproduce this center measured by \citet{Luger2017}, this does not necessarily mean that short-scale migration cannot reproduce the center, as the resonant centers are an artifact of initial conditions \citep[e.g.,][]{Tamayo2017}.

\begin{deluxetable}{rcccccc}
\tablecaption{Libration Angles for TRAPPIST-1\label{tab:trap1res}}
\tablecolumns{7}
\tablewidth{0.43\textwidth}
\tabletypesize{\footnotesize}
\tablehead{\colhead{} & \colhead{$\phi$} & \colhead{$\phi_c$} & \colhead{$\sigma_{\phi_c}$} & \colhead{$\phi_a$} & \colhead{$\sigma_{\phi_a}$}}   
\startdata
\textbf{lm}  & $\phi_1$  & \nodata & \nodata & \nodata & \nodata & 1  \\
             & $\phi_2$  & \nodata & \nodata & \nodata & \nodata & 1  \\
             & $\phi_3$  & 161.07 & 6.34 & 19.93 & 29.79 & 3  \\
             & $\phi_4$  & \nodata & \nodata & \nodata & \nodata & 1 \\
             
\textbf{sm}  & $\phi_1$  & 165.78 & 8.24 & 31.49 & 44.76 & 10  \\
             & $\phi_2$  & 30.65 & 16.39 & 15.50 & 17.81 & 9  \\
             & $\phi_3$  & 140.84 & 14.32 & 12.36 & 15.78 & 11  \\
             & $\phi_4$  & 183.00 & 11.78 & 94.81 & 2.68 & 5  \\
             
\textbf{ecc} & $\phi_1$  & 170.66 & 8.80 & 39.04 & 44.89 & 6  \\
             & $\phi_2$  & 27.29 & 15.48 & 15.35 & 17.07 & 13  \\
             & $\phi_3$  & 161.16 & 114.97 & 46.06 & 40.23 & 9  \\
             & $\phi_4$  & 182.61 & 16.99 & 95.14 & 2.38 & 6  
\enddata
\tablecomments{Resulting libration centers ($\phi_c$) and amplitudes ($\phi_a$) from our simulations, as well as the number of angles at each center. $\phi_1=2\lambda_b-5\lambda_c+3\lambda_d$, $\phi_2=\lambda_c-3\lambda_d+2\lambda_e$, 
$\phi_3=2\lambda_d-5\lambda_e+3\lambda_f$, and $\phi_4=\lambda_e-3\lambda_f+2\lambda_g$ denote the three-body angles between adjacent trios of planets. `lm', `sm' and `ecc' stand for long-scale migration, short-scale migration, and eccentricity damping, respectively. Angles are repeated when there are multiple libration centers. All parameters are in degrees and wrapped between [0,360]. See Table \ref{tab:K80res} for a description of variables. For $\phi_1$, $\phi_2$, and $\phi_4$ from the long-scale migration simulations, there was only one simulation in which the angle librated. Because of this, we do not include statistics from the simulations. For the three dynamical histories, the following number of simulations resulted in resonance: 11/300, 4/600, 5/300 for short-scale migration, long-scale migration, and eccentricity damping, respectively. Simulations that were in resonance but did not have their three-body angles librating had all two-body angles librating.}
\end{deluxetable}

All three dynamical histories were able to reproduce the observed resonant chain in TRAPPIST-1, eccentricities within the range estimated by \citet{Gillon2017}, and the observed period ratios.


\section{Conclusion}\label{sec:conclusion}

Given that migration is such a robust process, many have invoked it to explain the architectures of exoplanet systems, including the formation of resonant chains in specific systems. However, exoplanets display a broad range of period ratios, with only a few planets in or near resonance, and recent studies have shown the feasibility of in situ formation of close-in planets. These two factors bring into question how large of a role migration plays in sculpting the exoplanetary population. 
Using numerical simulations, we investigated whether long-scale migration is required to form the resonant chains seen in Kepler-80, Kepler-223, Kepler-60, and TRAPPIST-1 or if short-scale migration or eccentricity damping are plausible dynamical histories.  For long-scale migration, we assume the planets form at least 1AU from their host star and we migrate the planets in to their present-day locations by applying forces that reduce the semi-major axis and eccentricity of the outermost planet. 
For short-scale migration, we apply the same types of forces, although over a shorter distance, to planets just outside of their observed commensurabilities. For eccentricity damping, we start the planets just outside of their present-day locations, but only damp their eccentricities. These latter two dynamical histories are consistent with in situ formation of the planets. We find that we cannot conclude for any given system that long-scale migration is required, as we were able to reproduce the observed resonant chains in all three systems via all three possible histories.

The systems studied herein, as well as HR 8799 and GJ 876, are the only currently known exoplanetary systems containing confirmed resonant chains. With the launching of TESS and other future  exoplanetary missions, we might soon find more resonant chains that can be studied to see if formation mechanisms other than long-scale migration are possible. The formation of these systems and their architectures remains an open problem whose solution might shed some light on the role of migration in establishing system architectures.

In principle, we can favor a dynamical history based on how often it creates resonances with the observed properties (e.g., resonant center and amplitude). However, any results drawn from these simulations are dependent on the initial conditions where the planets form. Studies have found that centers and amplitudes are sensitive to initial conditions  \citep[e.g.,][]{Mustill2011,Tamayo2017}. A study in which a larger range of initial conditions are considered is warranted, but beyond this proof of concept exploring whether systems could have been put into resonance using these different dynamical histories.

Many simulations of long-scale migration resulted in the planets locking into the wrong resonance. This outcome was independent of how close the planets were initially to resonance.  Locking into the wrong resonance occurred for nearly all long-scale migration simulations for the Kepler-223 d/e pair; the pair is in a 4:3 orbital resonance, and yet nearly all of our simulations locked the planets into the 3:2 resonance. In a similar way, the TRAPPIST-1 b/c pair is suggested to be in an 8:5 orbital resonance \citep{Gillon2017}, yet in our simulations it locked into the 3:2 resonance as well as the 9:5 and the 7:4. These outcomes demonstrate that although long-scale migration is appealing for requiring less fine tuned initial conditions (i.e., the planets do not have to happen to form near resonance), planets can easily get trapped into the wrong resonance. This issue of trapping the planets in the incorrect resonance should be weighed against the less fine tuning in order to determine which dynamical history is more likely.


Analytic studies have shed some light on the formation of multi-body resonances \citep[e.g.,][]{Snellgrove2001,Papaloizou2015,Delisle2017}. However, the formation of resonant chains via eccentricity damping only from the gas disk has not, to our knowledge, been studied analytically. Once we understand analytically if there is any expected distinction between resonance capture through these mechanisms, we might be able to determine a preferred dynamical history. For example, if we were to find that a specific dynamical history always reproduces the measured libration center, we might be inclined to favor that history. Such conclusions would, however, require the amplitude and center for the resonant libration to be known very well, as is the case with Kepler-223 and Kepler-80.

If we can distinguish between the two types of multi-body resonances--one with two-body resonances librating and the other without the two-body libration
--then we can gain further insight into the dynamical history. When a multi-body resonance is formed via planets locking into two-body MMR, one after another, the two-body angles and the multi-body angle all librate \citep{Gozdziewski2016}. 
In our simulations, we found most of the three-body angles to be of the first type where the two-body angles also librated. However, some of simulations result in the systems being in the other type of resonance, 
but since the results of our simulations can change depending on the initial conditions, we need to first understand analytically if there is any distinction between resonance capture in the different dynamical histories.

\begin{acknowledgments}

We thank the referee for a helpful review that improved this paper. We thank Eric Agol for his feedback. MGM thanks Eric Ford, Bradford Foley, Jim Kasting, and Jason Wright for their helpful feedback. MGM acknowledges that this material is based upon work supported by the National Science Foundation Graduate Research Fellowship Program under Grant No. DGE1255832. Any opinions, findings, and conclusions or recommendations expressed in this material are those of the author and do not necessarily reflect the views of the National Science Foundation. RID gratefully acknowledges support from NASA XRP 80NSSC18K0355. The Center for Exoplanets and Habitable Worlds is supported by the Pennsylvania State University, the Eberly College of Science, and the Pennsylvania Space Grant Consortium.

\end{acknowledgments}

\bibliographystyle{apj}
\bibliography{all.bib}

\end{document}